\documentclass[10pt, titlepage, a4paper, times]{article}
	\usepackage{color,soul}
	\usepackage{dirtytalk}
    \usepackage{subfig}
    \usepackage{graphicx}
	\usepackage{amsmath}
	\usepackage{amsfonts}
	\usepackage{fancyhdr}
	\usepackage{enumerate}
	\usepackage{listings}
	\usepackage[titletoc]{appendix}
	\usepackage[pdfborder={0 0 0},colorlinks=true, urlcolor=blue, citecolor=red, bookmarks=false]{hyperref}
	\usepackage[margin=3cm]{geometry}
\usepackage[absolute]{textpos}
	\usepackage[section]{placeins}
	\usepackage{url}
	\usepackage{tabularx}
	\usepackage{caption}
	\usepackage[ruled,vlined]{algorithm2e}
	\usepackage{comment}
    \usepackage[dvipsnames]{xcolor}
	\renewcommand\contentsname{Table of Contents}

        \renewcommand{\thesection}{\thepart \arabic{section}.}
        \renewcommand{\thesubsection}{\thepart \arabic{section}.\arabic{subsection}.}

\makeatletter

\newcommand\frontmatter{%
    \cleardoublepage
  \pagenumbering{roman}}

\newcommand\mainmatter{%
    \cleardoublepage
  \pagenumbering{arabic}}

\newcommand\backmatter{%
  \if@openright
    \cleardoublepage
  \else
    \clearpage
  \fi
   }

\makeatother

\begin{document}
\begin{titlepage}
% Title
%\title{	

\begin{center}
		\begin{figure}[t]	
				\includegraphics[width=15mm, bb=0 0 100 100]{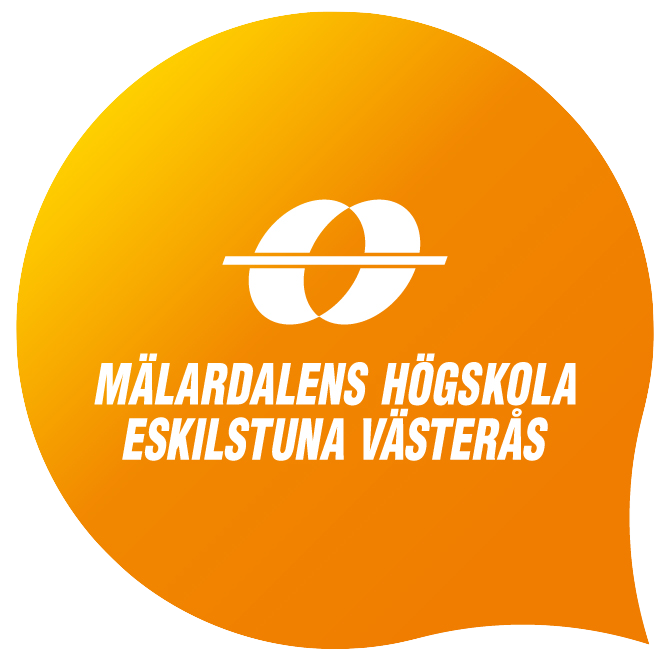}
		\end{figure}
                 	\Large M\"{a}lardalen University \\
			\Large School of Innovation Design and Engineering \\
                        \Large V\"{a}ster\r{a}s, Sweden\\

                        \noindent\makebox[\linewidth]{\rule{\textwidth}{0.4pt}}\\[0.5cm]
                
                %The complete name of the course you are enrolled in
                \Large{Thesis for the Degree of Master of Science in Computer Science with Specialization in Software
Engineering 30.0 credits}\\[2.0cm]

			\huge \textbf{\uppercase{Reinforcement Learning Assisted Load Test Generation for e-commerce applications}} \\ [2.5cm] %TITLE!!!!!!!!
				
			\LARGE Golrokh Hamidi   \\	
        	\large ghi19001@student.mdh.se \\[2.5cm]			

\begin{flushleft}
			\Large Examiner: \begin{minipage}[t]{0,7\textwidth}\Large Wasif Afzal\\\large M\"{a}lardalen University, \large V\"{a}ster\r{a}s, Sweden\\ \end{minipage}\\[0.5cm]
		%	\Large M\"{a}lardalen University\\
		%	\Large V\"{a}ster\r{a}s, Sweden\\[1.0cm]
			
			\Large Supervisors: \begin{minipage}[t]{0,7\textwidth}\Large Mahshid Helali Moghadam\\\large M\"{a}lardalen University, \large V\"{a}ster\r{a}s, Sweden \end{minipage} \\[0.5cm]
		%	\Large M\"{a}lardalen University\\
		%	\Large V\"{a}ster\r{a}s, Sweden\\[0.5cm]
			
                        \Large Company supervisor: \begin{minipage}[t]{0,6\textwidth}\Large Mehrdad Saadatmand, \\\large RISE - Research Institutes of Sweden, V\"{a}ster\r{a}s, Sweden \end{minipage}
		%	\Large Company\_Name\\
		%	\Large Location
\end{flushleft}

              \vspace*{\fill}
                    \large \today		%today should be replaced by the date the report is sent for examination

\end{center}
\end{titlepage}
%\date{}
%\maketitle

% Page style
\thispagestyle{fancy}
\fancyhead[R]{Reinforcement Learning Assisted Load Test Generation}
\fancyhead[L]{Golrokh Hamidi}
\fancyfoot[L]{}
%\fancyfoot[LE,RO]{\thepage}
\renewcommand{\headrulewidth}{0.4pt}
\renewcommand{\footrulewidth}{0.4pt}

% Begin actual text

\def\abstract{
  \vfil
\begin{center}%
{\bfseries\abstractname\vspace{-.5em}}
\end{center}
\itshape
}

\def\endabstract{\par
}

\frontmatter

% ============================= Abstract ==============================
\begin{abstract}

Background: End-user satisfaction is not only dependent on the correct functioning of the software systems but is also heavily dependent on how well those functions are performed. Therefore, performance testing plays a critical role in making sure that the system responsively performs the indented functionality. Load test generation is a crucial activity in performance testing. Existing approaches for load test generation require expertise in performance modeling, or they are dependent on the system model or the source code.

Aim: This thesis aims to propose and evaluate a model-free learning-based approach for load test generation, which doesn't require access to the system models or source code.

Method: In this thesis, we treated the problem of optimal load test generation as a reinforcement learning (RL) problem. We proposed two RL-based approaches using q-learning and deep q-network for load test generation. In addition, we demonstrated the applicability of our tester agents on a real-world software system. Finally, we conducted an experiment to compare the efficiency of our proposed approaches to a random load test generation approach and a baseline approach.

Results: Results from the experiment show that 
%the random load test generation approach performed better than the baseline approach (in terms of optimal workload generation). 
the RL-based approaches learned to generate the effective workloads with smaller sizes and in fewer steps. The proposed approaches led to higher efficiency than the random and baseline approaches.
%Results indicate that the DQN-based agent was able to generate workloads of optimal sizes that violate the performance properties.

Conclusion: Based on our findings, we conclude that RL-based agents can be used for load test generation, and they act more efficiently than the random and baseline approaches. 
%The hyper-parameters for DQN can also be tuned for better results in load test generation tasks.\\

\end{abstract}
\newpage

%========================== Table of contents ===========================
\hypersetup{linkcolor=black}
\tableofcontents
%\hypersetup{linkcolor=red}
\iffalse
\vfil
Inneh\r{a}llsf\"{o}rteckningen ska ange de olika rubrikerna i rapporten och p\r{a} vilka sidor i rapporten dessa finns. En f\"{o}rteckning \"{o}ver figurer kan finnas efter inneh\r{a}llsf\"{o}rteckningen. Samtliga figurer/bilder ska vara numrerade och refererade till i texten.
\fi
\clearpage

\mainmatter

% ============================== Content: Modify the structure according to your needs ===============================
\section{Introduction}
\label{sec:intro}
\paragraph{}
The industry is continuously finding ways to make software services accessible to more and more customers. One way to reach such customers (distributed over the globe) is the use of Enterprise Applications (EAs) delivering services over the internet. Inefficient and time-wasting software applications lead to customer dissatisfaction and financial losses~\cite{mm@web,glindenmakedatausefull}.
Performance problems are costly and waste resources. Furthermore, nowadays, using internet services and web-based applications have been extremely widespread among people and the industry. The significant role of internet services in people's daily life, and the industry is undeniable. Users around the world are dependent on internet services more than ever. Consequently, software success depends not only on the correct functioning of the software system but also on how well those functions are performed (non-functional properties). Responsiveness and efficiency are primitive requirements for any web-application due to the high expectations of users. For example, Google reported a 0.5 seconds increased delay in generating the search page resulted in 20\% decrease in traffic by users~\cite{mm@web}. Amazon also reported a 100 mili seconds delay in a web-page costed 1\% loss in sales~\cite{glindenmakedatausefull}. Accordingly, performance is a key success factor of software products, and it is of paramount importance to the industry, and a critical subject for user satisfaction. Tools allow companies to test software performance in both the development and design phases or even after the deployment phase.

%2. Introduction to performance testing\\
\paragraph{}
Performance describes how well the system accomplishes its functionality. Typically, the performance metrics of a system are response time, error rate, throughput, utilization, bandwidth, and data transmission time. 
Finding and resolving performance bottlenecks of a system is an important challenge during the development and maintenance of a software~\cite{jin2012understanding}.
The issues reported after project release are often performance degradation rather than system failures or incorrect response~\cite{weyuker2000experience}. 
Two common approaches to performance analysis are performance modeling and performance testing.
Performance models can be analyzed mathematically, or they could be simulated in case of having complex models~\cite{lavenberg1983computer}.
Measuring and evaluating performance metrics of the software system through executing the software under various conditions by simulating concurrent multi-users with tools is the core of performance testing.
One type of performance testing is load testing. Load testing evaluates the system's performance (e.g., response time, error rate, resource utilization) by applying extreme loads on the system~\cite{zhang2011automatic}. The load testing approaches usually generate workloads in multiple steps by increasing the workload in each step until a performance fault occurs in the system under test. The performance faults are triggered due to a higher error rate or response time than expected by the performance requirements \cite{zhang2011automatic}. Different approaches have been proposed for generating the test workload. 
%3. Introduction to RL based performance testing\\
%\paragraph{}
%talk about some approaches that require models and source code
Over the years, many approaches have been focused on testing for performance using system models or source code \cite{zhang2011automatic, zhang2002automated}. These approaches require expertise in performance modeling, and the source code of the system is not always available.
Various machine learning methods are also used in performance testing \cite{syer2011identifying,koo2019pyse}. However, these approaches require a significant amount of data for training. On the other hand, model-free Reinforcement Learning (RL)~\cite{sutton1998introduction} is one of the machine learning techniques which does not require any training data set. Unlike other machine learning approaches, RL can be used in load testing to generate effective\footnote{Effective, in terms of causing the violation of performance requirements (error rate and response time thresholds).
} workloads without any training data-set. 

%4. Related work\\
%\paragraph{}

%5. Limitations of existing work and the benefits of my approach\\
\paragraph{}
As mentioned before, in software systems, performance bottlenecks could cause violations of performance requirements~\cite{ibidunmoye2015performance, chandola2009anomaly}. Performance bottlenecks in the system will change during the time due to the changes in their source code. Load testing is a kind of performance testing in which the aim is to find the breaking points (performance bottlenecks) of the system by generating and applying workloads on the system. Manual approaches for test workload generations consume human resources; they are dependent on many uncontrolled manual factors and are highly prone to error. A possible solution to this problem is automated approaches for load testing. Existing automated approaches are dependent on the system model and may not be applicable when there is no access to the model or source code.
There is a need for a model-free approach for load testing, which is independent of source code, system model, and requires no training data.

\paragraph{Contributions.}
In this thesis, our purpose is to generate efficient\footnote{Efficient, in terms of optimal workload (workload size and number of steps for generating the workload).} workload-based test conditions for a system under test without access to source code or system models, based on using an intelligent RL load tester agent. Intelligent here means that the load tester tries to learn how to generate an efficient workload. The contributions of this thesis are as follows.
\begin{enumerate}
    \item Proposed model-free RL approach for load testing.
    \item An evaluation of the applicability of the proposed approach on a real case.
    \item An experiment for evaluating the two RL-based methods used in the approach i.e., q-learning and Deep Q-Network (DQN), against a baseline and a random approach for load test generation.
\end{enumerate}

%7. method
\paragraph{Method.}
In our proposed model-free RL approach, the intelligent agent can learn the optimal policy for generating test workloads to meet the performance analysis's intended objective. The learned policy might also be reused in further stages of the testing. In this approach, the workload is selected in an intelligent way in each step instead of just increasing the workload size. We explain our mapping of the real-world problem of load test generation into an RL problem. 
We also presented the RL methods that we use in our approach i.e., q-learning and deep q-network (DQN). Then we present our approach with two variations of RL methods in detail.
To evaluate the applicability of our proposed approach, we implement our RL-based approaches using open source libraries in Java. We use JMeter to generate our desired workload and apply the workload on an e-commerce website, deployed on a local server.

In addition, we conduct an experiment to evaluate the efficiency of the RL-based approaches. We execute the RL-based approaches, a baseline approach, and a random approach separately for comparison. We then compare the results of all approaches based on the efficiency (i.e., final workload size that violates the performance requirements and the number of workload increment steps for generating the workload).

%9. Results 10. Concluding para\\
\paragraph{Results} 
The experiment results show that, in comparison to the other approaches, the baseline approach generates workloads with bigger sizes. Thus the baseline approach is not as efficient as the other approaches. 
The random approach performs better than the baseline approach since
the average workload size generated by the random approach is lower than the baseline approach. However, the proposed RL-based approaches perform better than the random and baseline approaches.
The results show that in both q-learning and DQN approaches, efficient workload size and the number of steps taken for generating workload in each episode converges to a lower value over time. 
The q-learning approach converges faster than the DQN. However, the DQN approach converges to lower values for the workload sizes.
Our conclusion of the results is that both of the proposed RL approaches learn an optimal policy to generate optimal workloads efficiently.

\paragraph{Structure:}
    The remainder of this thesis is structured as follows. In Section \ref{sec:background}, we describe the basic knowledge and terms in performance testing and reinforcement learning.
    In Section \ref{sec:relatedwork}, we introduce different approaches for load testing.
    In Section \ref{problem}, we describe the motivation and problem, research goal, and research questions.
    In Section \ref{sec:method}, we present the scientific method we use in this thesis and the tools we used.
    In Section \ref{sec:approach}, we provide our approach for generating load test and explain our RL-based load testers in detail.
    In Section \ref{sec:evaluation}, we provide an overview of the SUT setup, the process of applying workload using JMeter, and the implementation of our load tester.
    In Section \ref{sec:results}, we describe the outcome of executing the implemented load testers on the SUT. We also explain the experiment procedure in this section.
    In Section \ref{sec:discussion}, we present an interpretation of the results.
    Finally, in Section \ref{sec:conclusions}, we summarize the thesis report and presents conclusions and future directions.

\newpage
\section{Background}
\label{sec:background}

%\instructions{In this section you describe the knowledge that the reader needs to understand your work and your contribution. Present basic knowledge needed to understand the area and the task. For example, you can describe relevant theories here and explain concepts you use or introduce mathematical notation. Write the background so that anyone who is well versed in the area can skip it.}
%\bigskip

In this section, we provide basic knowledge, terms, and notations in performance testing and reinforcement learning. The terms explained here will be used for describing the problem, approach, and solution in the following sections.

\subsection{Performance}

In this section, we discuss the terms related to performance and performance testing.

\paragraph{ Non-functional Quality Attributes of Software}
Non-functional properties of a software system define the physiognomy of the system. These non-functional properties are often achieved by realizing some constraints over the functional requirements. Performance, security, availability, usability, interoperability, etc., are often classified under the term of run-time non-functional requirements. Then, modifiability, portability, reusability, integrability, testability, etc., are considered as non-runtime non-functional requirements. The run-time non-functional requirements can be verified by performance modeling in the development phase or by performance testing in execution.

\paragraph{Performance}
Performance is of paramount importance in connected systems and is a key success factor of software products. For example, EAs [1] such as e-commerce providing services to the customer over the globe, their success is subjected to the performance.
Performance describes how well the system accomplishes its functionality. Efﬁciency is another term that is used in place of performance in some classiﬁcations of quality attributes \cite{ISO/IEC,glinz2007non,chung2012non}. Some performance metrics or performance indicators are:
\begin{itemize}
\item Response Time: The time between sending a request and beginning to receive the response.
\item Error Rate: The proportion of erroneous units of transmitted data.
\item Throughput: The number of processes that a system can handle per second.
\item Utilization on computer resources: e.g., processor usage and memory usage.
\item Bandwidth: The maximum rate of the data transferred in a given amount of time.
\item Data Transmission Time: The amount of time that it takes for the transmitting node to put all the data on the wire.
\end{itemize}
Performance is one of the important factors that should also be taken into consideration in the design, development, and configuration phase of a system \cite{lavenberg1983computer}.

\subsubsection{Performance Analysis}
The performance of a system could be evaluated through measurements manually in a user environment or under controlled benchmark conditions \cite{lavenberg1983computer}.
Two conventional approaches to performance analysis are performance modeling and performance testing.

\paragraph{Performance Modeling}
It is not always feasible to measure the performance of the system or component, for example, in the design and development phase. In this case, the performance could be predicted based on models. Performance modeling is used during the design and development, and for configuration tuning and capacity planning. Other than quantitative predictions, performance modeling will give us insight into the structure and behavior of the system during the system design. To acquire performance measures, performance models can be analyzed mathematically, or they can be simulated in case of having complex models \cite{lavenberg1983computer}.
Some of the well-known modeling notations are queuing networks, Markov processes, and Petri nets, which are used together with analysis techniques to address performance modeling.
\cite{cortellessa2011model,harchol2013performance,kant1992introduction}.

\iffalse
\subparagraph{Queuing Networks}
\subparagraph{Markov Processes}
\subparagraph{Perti Nets}
Petri nets (PN) are similar to Finite State Automata. The difference is, in Petri nets, each state is a set of partial and independent states of automata, and a transaction considers a partial state of the automata, not the global state of the system \cite{cortellessa2011model}. Petri nets are a formal modeling technique to specify the synchronization behavior of concurrent systems \cite{cortellessa2011model}."
\fi

\paragraph{Performance Testing}

%\todoB{update this sectiond:https://www.linkedin.com/pulse/performance-load-stress-endurance-test-which-do-you-want-chris-jones}
The IEEE standard definition of performance testing is: “Testing conducted to evaluate the compliance of a system or component with specified performance requirements”~\cite{10.5555/574566}.
Measuring and evaluating the response time, error rate, throughput, and other performance metrics of the software system through executing the software under various conditions by simulating concurrent multi-users with tools is the core of performance testing. Performance testing could be performed on the whole system or on some parts of the system. Performance testing can also validate the efficiency of the system architecture, the system configurations, and the algorithms used by the software \cite{jiang2015survey}. Some types of performance testing are load testing, stress testing, endurance testing, spike testing, volume testing, and scalability testing.

\paragraph{Performance Bottlenecks}
Performance bottlenecks will result in violating performance requirements \cite{ibidunmoye2015performance, chandola2009anomaly}. The definition of a performance bottleneck is any system, component, or a resource that restricts the performance and prevents the whole system from operating properly as required \cite{gregg2013systems}. The source of performance anomalies and bottlenecks are \cite{ibidunmoye2015performance}:
\begin{itemize}
\item Application Issues: Issues in the application-level like incorrect tuning, buggy codes, software updates, and incorrect application conﬁguration 
\item Workload: Application loads can effect in congested queues and resource and performance issues. 
\item Architectures and Platforms: For example, the behavior and effects of the garbage collector, the location of the memory and the processor, etc. can affect the system's performance.
\item System Faults: Faults in system resources and components such as software bugs, operator error, hardware faults, environmental issues, and security violations.
\end{itemize}

\paragraph{Load Testing}
The load is the rate of different requests that are submitted to a system \cite{beizer1984software}. Load testing is the process of applying load on software to observe the software behavior and detect issues caused because of the load \cite{jiang2015survey}. Load testing is applied through simulating multiple users to access the software at the same time.

\paragraph{Regression testing}
Testing the software after new changes in the software is called regression testing. The aim of regression testing is to ensure the previous functionality of the software has not been violated, and it still meets the functional and non-functional requirements.

%\subsubsection{Performance Testing Process}
%\todoB{https://www.guru99.com/performance-testing.html}

\paragraph{Performance Testing Tools}
There are a variety of Performance Testing tools for measuring web application performance and load stress capacity. Some of these tools are open-source, and some have free trials. Some of the most popular performance testing tools are Apache JMeter, LoadNinja, WebLOAD, LoadUI, LoadView, NeoLoad, LoadRunner, etc.

\subsection{Machine Learning}\label{machine_learning}
Nowadays, Machine Learning plays an important role in software engineering and is widely used in computer technology. Some well-known applications of machine learning algorithms in software engineering are:
\begin{itemize}
    \item Test data generation: Transforming speech to text.
    \item Drive autonomous vehicles: For example, google self-driving cars.
    \item Image Recognition: Detecting an object in a digital image.
    \item Sentiment Analysis: Determining the attitude or opinion of the speaker or the writer.
    \item Prediction: For example, traffic prediction and weather prediction.
    \item Information Extraction: Extracting information from unstructured data.
    \item Medical diagnoses: Medical diagnoses based on clinical parameters.
\end{itemize}
Machine learning algorithms are a set of methods and algorithms in which the computer program learns to improve a task with respect to a performance measure based on experience. Machine learning uses techniques and ideas from artificial intelligence, probability and statistics, computational complexity theory, control theory, information theory, philosophy, psychology, neurobiology, and other fields \cite{10.5555/541177}. 
Three major categories in learning problems are:
\begin{itemize}
\item Supervised Learning
\item Unsupervised Learning
\item Reinforcement Learning
\end{itemize}

\begin{figure}[h]
\centering
\includegraphics[width=0.6\linewidth]{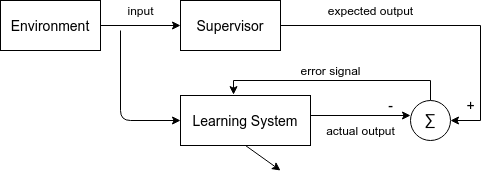}
\caption{Supervised Learning}\label{sl}
\end{figure}

\paragraph{Supervised Learning}
In supervised learning, the training data set provides an output variable corresponding to each input variable. Supervised learning predicts the classification of other unlabeled data in the test data set based on the labeled training data in the training data set. Regression and classification are two types of supervised learning. The target is to minimize the expected output and the actual output of the learning system. Figure \ref{sl}

\paragraph{Unsupervised Learning}
In unsupervised learning, unlike supervised learning, The training data set does not contain the output value of each input set, i.e., the training data set is not labeled. Unsupervised learning algorithms take unlabelled data as input and cluster the data in the same group based on their attributes.

\subsubsection{Reinforcement Learning}\label{sec:background_RL}
\begin{figure}[h]
\centering
\includegraphics[width=0.6\linewidth]{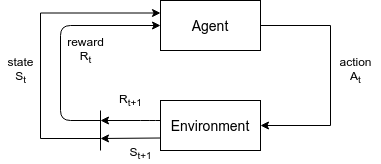}
\caption{Reinforcement Learning}\label{rl}
\end{figure}

In reinforcement learning, the agent tries to learn the best policy by experimenting and trial and error interaction with the environment. Reinforcement learning is goal-directed learning in which the goal of the agent is to maximize the reward \cite{10.5555/541177}. In reinforcement learning problems, there is no training data set. In this case, the agent itself explores the environment to collect data and update its policy to maximizes its expected cumulative reward over time (illustrated in Figure \ref{rl} \cite{sutton1998introduction}). "Trial-and-error search and delayed reward are the two most important distinguishing features of reinforcement learning"\cite{sutton1998introduction}. The agent is the learner and decision-maker, and everything outside the agent is the environment. The state is the current situation that is returned by the environment. Each action results in a new state and gives a reward corresponding to the state (or state-action). In a reinforcement learning problem, the reward function specifies the goal of the problem \cite{sutton1998introduction}. It is not specified for the agent which action to take in each state, and instead, the agent should discover taking which action leads to the most reward by trying them. 
% write down section 3.1

In the following, we discuss the main concepts in reinforcement learning:\\

\paragraph{Agent and Environment}
Everything except the agent is the environment; everything that the agent can interact with directly or indirectly. When the agent performs actions, the environment changes. This change is called the state-transition. As shown in Figure \ref{rl} \cite{sutton1998introduction}, At each step $t$, the agent executes action $A_t$, receives a representation of the environments state $S_t$ based on the observations from the environment, and receives a reward $R_t$.

\paragraph{State}
State contains the information used to determine what happens next. History is a sequence of states, actions, and rewards:
\begin{equation}
\label{eq:history}
    H_t = S_0, A_0, R_1, ...,S_t, A_t, R_{t+1}
\end{equation}
The agent state is a function of the history:
\begin{equation}
\label{eq:state}
    S_t = f(H_t)
\end{equation}

\paragraph{Action}
Actions are the agent's decision, which leads to a next state and provides a reward from the environment. Actions affect the immediate reward and can also affect the next state of the agent and consequently, the future rewards (delayed reward). So the actions may have long term consequences. The policy determines which action should be taken in each step.

\paragraph{Reward and Return}
 A reward $R_t$ is a scalar feedback signal that shows how well the agent is operating at step $t$. The learning agent tries to reach the goal of maximizing cumulative reward in the future. The Reward may be delayed, and it may be better to sacrifice immediate reward to gain more long-term reward. Reinforcement learning is based on the reward hypothesis: "All goals can be described by the maximization of expected cumulative reward". $R_s^a$ shown in equation \ref{eq:reward_function_equation} \cite{sutton1998introduction} is the expected value of the reward of taking action $a$ from state $s$.

\begin{equation}
\label{eq:reward_function_equation}
    R_s^a = \mathbb{E}[R_{t+1}|S_t=s, A_t=a]
\end{equation}
The return $G_t$ in equation \ref{eq:return} \cite{sutton1998introduction} is the total discounted reward from time-step $t$.

\begin{equation}
    \label{eq:return}
    G_t = R_{t+1}+\gamma R_{t+1}+... = \sum_{k=0}^{\infty} \gamma^{k}R_{t+k+1}
\end{equation}

\paragraph{Discount Factor}
Discount factor $\gamma$ is a value in the interval (0,1]. A reward that occurs $k+1$ steps in the future is multiplied by $\gamma^k$, which means the value of receiving reward $R$ after $k+1$ time-steps is decreased to $\gamma^{k}R$. The discount factor indicates how much do we value the future rewards. The more we trust our model, the discount factor would be nearer to 1, but if we are not certain about our model, the discount factor would be near to 0.

\paragraph{Markov Decision Process (MDP)}
An MDP is an environment represented by a tuple \(\langle{S,A,P,R,\gamma} \rangle\). Where \textit{S} is a countable set of states, \textit{A} is a countable set of actions, \textit{P} is the state-transition probability function in equation \ref{eq:probability_function} \cite{sutton1998introduction}, \textit{R} is the reward function in equation \ref{eq:reward_function_equation}, and $\gamma$ is the discount factor \cite{sutton1998introduction}. The state-transition probability \(P_{ss'}^a\) is the probability of going to state $s'$ by taking the action $a$ from state $s$. Almost all reinforcement learning problems can be formalised as MDPs.

\begin{equation}
\label{eq:probability_function}
    P_{ss'}^a = \mathbb{P}[S_{t+1}=s'|S_t=s, A_t=a]
\end{equation}
In a MDP, the state is fully observable i.e the current state completely characterises the process. A state $S_t$ (current state in time $t$) is Markov if and only if it follows the rule in equation \ref{eq:markov_property} \cite{sutton1998introduction},

\begin{equation}
\label{eq:markov_property}
   \mathbb{P}[S_{t+1}|S_t]= \mathbb{P}[S_{t+1}|S_1, ..., S_t]
\end{equation}
meaning that the future state is only dependent of the present and it is independent of the past. A Markov state contains every relevant information from the history. So when the state is specified, the history may be thrown away.

\paragraph{Partially Observable Markov Decision Process (POMDP)}
In POMDP, the agent is not able to directly observe the environment, meaning the environment is partially observable to the agent. So unlike MDP, the agent state is not equal to the environment state. In this case, the agent must construct its own state representation.

\paragraph{Policy}
The policy $\pi$ is the agent's behavior function. It is a function from a state to action. A deterministic policy specifies which action should be taken in each state; it takes a state as an input and it's output is an action:
\begin{equation}
\label{eq:deterministic_policy}
   a = \pi(s)
\end{equation}
A stochastic policy (equation \ref{eq:stochastic_policy} \cite{sutton1998introduction}) determines the probability of the agent taking a specific action in a specific state:
\begin{equation}
\label{eq:stochastic_policy}
   \pi(a|s) = \mathbb{P}[A_t=a|S_t=s]
\end{equation}

\paragraph{Value function}
The value function is a prediction of future reward that is used to evaluate how good a state is. The value function $v_\pi(s)$ of a state $s$ under policy $\pi$ is the expected return of following policy $\pi$ starting from the state $s$. The value function for MDPs is shown in equation \ref{eq:state_value_function} \cite{sutton1998introduction}:
\begin{equation}
    \label{eq:state_value_function}
    v_\pi(s) = \mathbb{E}[G_t|S_t=s] = \mathbb{E}[\sum_{k=0}^{\infty} \gamma^{k}R_{t+k+1}|S_t=s]
\end{equation}
$v_\pi(s)$ is called the \textit{state-value function for policy $\pi$}. If terminal states exist in the environment, there value is zero.\

The value of taking action $a$ in state $s$ under policy $\pi$ is $q_\pi(s,a)$ which is called the \textit{action-value function for policy $\pi$} or the \textit{q-function} shown in equation \ref{eq:action_value_function} \cite{sutton1998introduction}:

\begin{equation}
    \label{eq:action_value_function}
    q_\pi(s,a) = \mathbb{E}[G_t|S_t=s, A_t=a] = \mathbb{E}[\sum_{k=0}^{\infty} \gamma^{k}R_{t+k+1}|S_t=s, A_t=a]
\end{equation}
$q_\pi(s,a)$ is the expected return starting from state $s$, taking the action $a$ and future actions based on policy $\pi$.

\begin{figure}[h]
\centering
\includegraphics[scale=0.5]{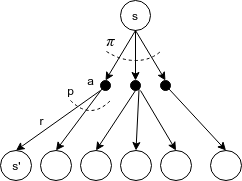}
\caption{Backup diagram for $v_\pi$}\label{backup_diagram_v}
\end{figure}

\paragraph{Bellman Equation}
The Bellman equation explains the relation between the value of a state or state-action with it's successors. The Bellman equation for $v_{\pi}$ is shown in equation \ref{eq:bellman_equation_v} \cite{sutton1998introduction}: 

\begin{equation}
    \label{eq:bellman_equation_v}
    \begin{split}
        v_\pi(s) & = \mathbb{E_\pi}[G_t|S_t=s] \\
        & = \mathbb{E_\pi}[R_{t+1} + \gamma G_{t+1}|S_t=s] \\
        & = \sum_{a}\pi(a|s)\sum_{s'}\sum_{r}p(s',r|s,a)\big[r+\gamma\mathbb{E}_\pi[G_{t+1}|S_{t+1}=s']\big] \\
        & = \sum_{a}\pi(a|s)\sum_{s',r}p(s',r|s,a)\big[r+\gamma v_\pi(s')\big]
    \end{split}
\end{equation}
Where $p(s',r|s,a)$ is the probability of going to state $s'$ and receiving the reward $r$ by taking the action $a$ from state $s$. Figure \ref{backup_diagram_v} \cite{sutton1998introduction} helps explaining the equation. Based on this equation, the value of a state is the average of it's successor states' values plus the reward of reaching them, weighting each state value by the probability of its occurrence.
This recursive relation of states is a fundamental property value function in reinforcement learning.

\noindent The Bellman equation for q-value (action value) $q_\pi(s,a)$ is shown in equation \ref{eq:bellman_equation_q} \cite{sutton1998introduction}: 

\begin{equation}
    \label{eq:bellman_equation_q}
        q_\pi(s,a) = \sum_{s',r}p(s',r|s,a)\big[r +\gamma\sum_{a'}\pi(a'|s')q_\pi(s',a')]
\end{equation}

\begin{figure}[h!]
\centering
\includegraphics[scale=0.5]{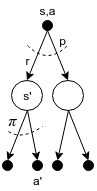}
\caption{Backup diagram for $q_\pi$}\label{backup_diagram_q}
\end{figure}

\noindent This equation is clarified in Figure \ref{backup_diagram_q} \cite{sutton1998introduction}.

\begin{figure}[h!]
\centering
\includegraphics[scale=0.5]{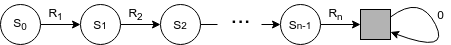}
\caption{Episode}\label{episode}
\end{figure}

\paragraph{Episode}
A sequence of states starting from an initial state and finishing in a terminal state is named episode. Different episodes are independent from each other. Figure \ref{episode} gives an overview of an episode.

\paragraph{Episodic and Continuous tasks}
There are two kinds of tasks in reinforcement learning; episodic and continuous. Unlike continuous tasks, episodic tasks are when the interaction of the agent with the environment is broken down into separate episodes.

\paragraph{Policy Iteration}
Policy iteration is the process of achieving the goal of the agent, which is finding the optimal policy $\pi^*$. Policy iteration consists of two parts; policy evaluation and policy iteration, which are executed iteratively. 
Policy evaluation is the iterative computation of the value functions for a given policy while the agent interacts with the environment. And policy improvement is enhancing the policy by choosing actions greedily with respect to the recently updated value function:

\begin{equation}
    \label{eq:policy_iteration}
    \pi_0 \xrightarrow{E} v_{\pi_0} \xrightarrow{I} \pi_1 \xrightarrow{E} v_{\pi_1} \xrightarrow{I} ...
     \xrightarrow{I} \pi_* \xrightarrow{E} v_{\pi_*} 
\end{equation}

\paragraph{Value Iteration}
Value iteration is finding optimal value function iteratively. When the value function is optimal, then the policy out of it is also optimal. Unlike policy iteration, there is no explicit policy in value iteration, and the actions are chosen directly based on the optimal (converged) value function. Finding optimal value function is a combination of policy improvement and truncated policy evaluation.

\paragraph{Exploration and Exploitation}
The reinforcement learning agent should choose the actions that have tried before, which have the highest return; this is exploitation. On the other hand, the agent should try new actions that have not selected before to find these best actions; this is exploration. There is a trade-off between exploration and exploitation in the learning process and making a balance between them is one of the challenges in reinforcement learning problems.

\paragraph{$\varepsilon$-Greedy Policy}
An $\varepsilon$-greedy policy allows performing both exploration and exploitation during the learning. $\varepsilon$ is a number in the range of [0,1] is chosen. In each step, the probability of selecting the best action (best action based on the main policy which is extracted from the q-table) is 1-$\varepsilon$, and a random action is selected by the probability of $\varepsilon$. 

\paragraph{Monte Carlo}
Monte Carlo methods are a class of algorithms that repeat random sampling to achieve a result. One of the methods used in reinforcement learning is the Monte Carlo method to estimate value functions and find the optimal policy by averaging the returns from sample episodes. In this method, each episodic task is considered as an experience, which is a sample sequence of states, actions, and rewards. By using this method, we only need a model that generates sample transitions, and there is no need for the model to have complete probability distributions of all possible transitions and rewards. A simple Monte Carlo update rule is shown in equation \ref{eq:MC_value_fucntion_update} \cite{sutton1998introduction}:
\begin{equation}
    \label{eq:MC_value_fucntion_update}
    V(S_t) \xleftarrow{} V(S_t)+ \alpha[G_t-V(S_t)]
\end{equation}
Where $G_t$ is the return starting from time $t$ and $\alpha$ is the step-size (learning rate).

\paragraph{Temporal-Difference (TD) learning}
Temporal-difference learning is another learning method in reinforcement learning. The TD method is an alternative to the Monte Carlo method for updating the estimation of the value function. The update rule for the value function is shown in equation \ref{eq:TD_value_fucntion_update} \cite{sutton1998introduction}:
\begin{equation}
    \label{eq:TD_value_fucntion_update}
    V(S_t) \xleftarrow{} V(S_t)+ \alpha[R_{t+1} + \gamma V(S_{t+1})-V(S_t)]
\end{equation}
Unlike Monte Carlo, TD learns from incomplete episodes. TD can learn after each step and does not need to wait for the end of the episode. The algorithm \ref{Algorithm: TD(0)} explains $TD(0)$ \cite{sutton1998introduction}:\\

\begin{algorithm}[H]
\SetAlgoLined
\caption{TD(0) for estimating $v_\pi$}
\label{Algorithm: TD(0)}
\textrm{Input: the policy $\pi$ to be evaluated}\\
\textrm{Algorithm parameter: step size $\alpha \in (0,1]$}\\
\textrm{Initialise $V(s)$, for all $s \in$ state space, arbitrarily except that \textit{V(terminal)} = 0}\\

\For{each episode}{
    Initialize $S$\\
    \For{each step of episode}{
        \textrm{$A \xleftarrow{}$ action given by $\pi$ for $S$}\\
        Take action $A$, observe $R,S'$\\
        $V(S) \xleftarrow{} V(S)+\alpha[R+\gamma V(S')-V(S)]$\\
        $S \xleftarrow{} S'$\\
    }
    until $S$ is terminal
}
\end{algorithm}

\paragraph{Experience Replay}

In a reinforcement learning algorithm, the RL agent interacts with the environment and updates the policy, value functions, or model parameters iteratively based on the observed experiment in each step. The data collected from the environment would be used once for updating the parameters, but it would be discarded in the future steps. This approach is wasteful because some experiences may be rare but useful in the future. Lin \textit{et al.} \cite{lin1992self} introduced experience replay as a solution to this problem. 
An experience (state-transition) in their definition \cite{lin1992self} is a tuple of (\textit{x, a, y, r}) which means taking action \textit{a} from state \textit{x} and going to state \textit{y} and getting the reward \textit{r}.
In the experience replay method, a buffered window of N experiences is saved in the memory, and the parameters are updated with a batch of transitions in the experience replay, which are chosen based on different approaches e.g., randomly \cite{mnih2015human} or prioritized experiences \cite{schaul2015prioritized}.
Experience replay allows the agent to reuse the past experiences in an effective way and use them in more than one single update as if the agent experiences what it has experienced before again and again. Experience replay will speed up the learning of the agent, which leads to quicker convergence of the network. In addition, faster learning leads to less damage to the agent (the damage is when the agent takes actions based on bad experiences; therefore, it experiences a bad experience again and so on). Experience replay consumes more computing power and more memory but reduces the number of experiments for learning and the interaction of the agent with the environment, which is more expensive. Schau \textit{et al.} \cite{schaul2015prioritized} explain many stochastic gradient-based algorithms which have the i.i.d. assumption which is violated by strongly correlated updates in the RL algorithm and experience replay will break this temporal correlation by applying recent and former experiences in each update. 
Using experience replay has been effective in practice, for example Mnih \textit{et al.} \cite{mnih2015human} applied experience replay in the DQN algorithm to stabilize the value function's training. Google DeepMind also significantly improved the performance of the "Atari" game by using experience replay with DQN.\\

%\paragraph{\todoB{Example}}

\subsubsection{Q-learning}

Q-learning is one of the basic reinforcement learning algorithms. Q-learning is an off-policy TD control algorithm. Methods in this family learn an approximator q-function for the optimal action-value function $Q^*$.
In this algorithm the q-values of every possible state-action pairs are stored in a table named q-table. The q-table is updated based on the Bellman equation \ref{eq:q_learning_update} \cite{sutton1998introduction}:

\begin{equation}
    \label{eq:q_learning_update}
    Q(S_t,A_t) \xleftarrow{} Q(S_t,A_t)+ \alpha[R_{t+1} + \gamma \max\limits_{a}Q(S_{t+1},a) -Q(S_t,A_t)]
\end{equation}\\
The action is usually selected by an $\varepsilon$-greedy policy. But the q-value is updated independent of the policy being followed (off-policy) algorithm, and based on the next action which has the maximum q-value. The q-learning algorithm is shown in Algorithm \ref{Algorithm: Q-learning)}.

\subsubsection{Deep RL}
Deep reinforcement learning refers to the combination of RL with deep learning. Deep RL
is nonlinear function approximation methods like artificial neural network (ANN) using SGD \cite{sutton1998introduction}.

\paragraph{Value Function Approximation}
 Function approximation is used in RL because in large environments there are too many states and actions to be stored in the memory, also it is too slow to learn the value of each state/state-action individually. So the idea is to generalize from the visited states to the states which have not been visited yet. Hence the value function is estimated with function approximation:
\begin{gather*} 
    \label{eq:fucntion_approximation}
    \hat{v}(s,w) \approx v_\pi(s)\\
    \hat{q}(s,a,w) \approx q_\pi(s,a)
\end{gather*}
Where $w$ is the weight vector, for example, $w$ is the feature weights in q linear function approximator, which returns the estimated value of each state by multiplying in the state's feature vector. The dimensionality of $w$ is much less than the number of states and changing the weight vector, changes the estimated value of many states, therefor when $w$ is updated after each action from a single state, not only the value of that specific state will update, many states' values will be updated too. This generalization makes learning faster and more powerful. Moreover, using function approximation makes reinforcement learning applicable to problems with partially observable environments. 

\begin{figure}[h!]
\label{value_functions_approximation_types}
\centering
\includegraphics[scale=0.7]{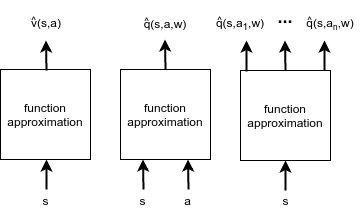}
\caption{Types of value function approximation}
\end{figure}

There are many function approximators, e.g., linear function of features, artificial neural network, decision tree, nearest neighbor, Fourier/wavelet bases, and etc. For value function, approximation differentiable function approximators are used e.g., linear function and neural networks.

\paragraph{Stochastic Gradient Descent}
Stochastic Gradient Descent or SGD is an optimization algorithm. This algorithm is used in machine learning algorithms, like training artificial neural networks used in deep learning. In this method, the goal is to find some model parameters which optimize an objective function by updating a model iteratively over multiple discrete steps. Optimizing an objective function is minimizing a loss function or maximizing a reward function (fitness function). In each step, the model makes some predictions based on the samples in the training data set, and based on the set of current internal parameters; then the predictions are compared to the real expected outcomes in the data set by calculating performance measures like mean square error. Then the gradient of the error is calculated and used to update the internal model's parameters to decrease the error. Sample size, batch size, and epoch size are some hyperparameters in SGD \cite{DifBatch}: 
\begin{itemize}
    \item Sample: A training data set contains many samples. A sample could be referred to as an instance, observation, input vector, or a feature vector. A sample is a set of inputs and an output. The inputs are fed into the algorithm, and the output is compared to the prediction by calculating the error.
    \item Batch: The model's internal parameters would get updated after applying a batch of samples to the model. At the end of applying each batch of samples to the model, the error is computed. The batch size can be equal to the training data set size (Batch Gradient Descent), it can be equal to 1 meaning each batch is a sample in the data set (Stochastic Gradient Descent), and it can be between 1 and the training set size (Mini-Batch Gradient Descent). 32, 64, and 128 are popular batch sizes in mini-batch gradient descent.
    \item Epoch: The whole training data set is fed to the model once in each epoch. In every epoch, each sample will update the internal model parameters for one time. So in an SGD algorithm, there are two for-loops; the outer loop is over the number of epochs, and the inner loop iterates over the batches in each epoch.
\end{itemize}
There is no specific rule for configuring these parameters. The best configuration differs for each problem and is obtained by testing different values.

\paragraph{Deep Q-Network (DQN)}
Deep Q-Network is a more complex version of q-learning. In this version, instead of using the q-table for accessing q-values, the q-values are approximated using an ANN.

\begin{figure}[h]
\centering
\includegraphics[width=0.6\linewidth]{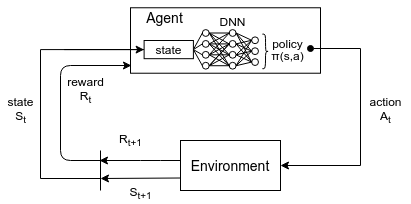}
\caption{Deep Q-Network}\label{dqn}
\end{figure}

%\subsubsection{Kinds of RL algorithms}
%\textcolor{blue}{\url{https://spinningup.openai.com/en/latest/spinningup/rl_intro2.html#part-2-kinds-of-rl-algorithms}}

\paragraph{Double Q-learning}
Simple q-learning has a positive bias in estimating the q-values; it can overestimate q-values. Double q-learning is an extension of q-learning which overcomes this problem. It uses two q-functions, and in each update, one of the q-functions is updated based on the next state's q-value from the other q-function \cite{hasselt2010double}. The double q-learning algorithm is shown in Algorithm \ref{Algorithm: double Q-learning)} \cite{hasselt2010double}:\\

\begin{algorithm}[H]
\SetAlgoLined
\caption{Double Q-learning}
\label{Algorithm: double Q-learning)}
\textrm{Initialize $Q^A, Q^B, s$}\\
\Repeat{end}{
    Choose $a$, based on $Q^A(s,.)$ and $Q^B(s,.)$, observe $r,s'$\\
    Choose (e.g. random) either UPDATE(A) or UPDATE(B)\\
    \If{UPDATE(A)}{
        Define $a* = \max\limits_{a} Q^A(s',a)$\\
        $Q^A(s,a) \xleftarrow{} Q^A(s,a)+ \alpha[r + \gamma Q^B(s',a^*)-Q^A(s,a)]$\\ 
        }
    \ElseIf{UPDATE(B)}{
        Define $b* = \max\limits_{a} Q^B(s',a)$\\
        $Q^B(s,a) \xleftarrow{} Q^B(s,a)+ \alpha[r + \gamma Q^A(s',b^*)-Q^B(s,a)]$\\ 
        }
    $s \xleftarrow{} s'$\\
}
\end{algorithm}
\paragraph{Double Deep Q Networks(DDQN)}
The idea of double q-learning can be used in DQN \cite{van2016deep}. There is an online network, a target network, and the online network gets updated based on the q-value from the target network. The target network is freezed and gets updated from the online network after N steps. The other way is to smoothly average for every N number of last updates. N is the "target DQN update frequency".

%\subsection{\todoB{Evolutionary Computation}}
 %SOTA
\newpage
\section{Related Work}
\label{sec:relatedwork}

%\instructions{The purpose of this section is to place your work in a context and compare it with previously published work and results in the field. This part should be thorough. You describe here existing knowledge and how this is expanded by your work. It should include analyzes of previous work that describe, for example, how different methods differ. You must show the most important similarities and differences regarding task, approach / methodology and results. It is important that you discuss in a neutral way the advantages and disadvantages of your own work compared to others.}

%\instructions{This also creates an expectation of the contribution for your work, the reader here learns about the limitations of previous work and why your task is a challenge.}

%\instructions{Together, this section, together with the background, will introduce "state of the art" / "state of practice" and its shortcomings, the importance of the task and what your work should be compared to.}\\

%In this section, we introduce the existing approaches for load testing. 
%We compare our reinforcement learning approach with the previous approaches and explain how does our approach differ from the existing ones and what are the advantages of our approach.\\

As mentioned before, in this study, we aim to detect certain workloads that cause performance issues in the software. To accomplish this objective, we use a reinforcement learning approach that applies workloads on the system and learns how to generate efficient workloads by measuring the performance metrics. Measuring performance metrics (e.g.,  response time,  error rate,  resource utilization) by applying various loads on the system under different execution conditions and different platform configurations is a common approach in performance testing \cite{ menasce2002load, apte2017autoperf, jindal2019performance}. Also discovering performance problems like performance degradation and violation of performance requirements that appear under specific workloads or resource configurations is a usual task in different types of performance testing \cite{briand2005stress, zhang2011automatic, ayala2018one}.\\

%similarity: \textbf{Measuring performance metrics} under different execution conditions including \textbf{various workload} and platform configurations \cite{ menasce2002load, apte2017autoperf, jindal2019performance}, [[30-32]]\\

%detecting different performance-related issues such as functional problems or violations of performance requirements which \textbf{emerge under certain workload} or resource configuration conditions \cite{briand2005stress, zhang2011automatic, ayala2018one}, [[33-35]] are often common \textbf{objectives of different types of performance testing.}\\

Different methods have been introduced for load test generation, e.g., analyzing system model, analyzing source code,  modeling real usage, declarative methods, and machine learning-assisted methods. We provide a brief overview of these approaches in the following:
\paragraph{Analyzing system model}
%Analysis of a performance model of SUT in terms of Petri nets using constraint solving techniques \cite{zhang2002automated}, \
Zhang and Cheun \cite{zhang2002automated} introduce an automatable method for stress test generation in terms of Petri nets.
%using genetic algorithms to generate load test based on the control flow graph of SUT \cite{gu2009search, di2007search}, \
Gu and Ge \cite{gu2009search} use genetic algorithms to generate performance test cases, based on a usage pattern model from the system’s workflow.
%a method to generate performance test cases automatically based on Genetic Algorithms for any system consisting of composite services. It considers users’ experience in the performance test model. Based on the system’s workflow, the usage pattern of a system is modeled\
Again Penta \textit{et. al.} \cite{di2007search} generate test data with genetic algorithms using workflow models. 
 %applying genetic-based algorithms to other types of system models such as UML models to generate stress load test \cite{garousi2010genetic, garousi2008traffic, costa2012generating, da2011generation} are samples of the techniques proposed in this category.\
Garousi \cite{garousi2010genetic} provides a genetic algorithm based UML-driven tool for stress test requirements generation.
Again Garousi \textit{et. al.} \cite{garousi2008traffic} introduce a UML model-driven stress test method for detecting network traffic anomalies in distributed real-time systems using genetic algorithms\
%\cite{costa2012generating} not written yet
%\cite{ da2011generation} not written

\paragraph{Analyzing source code}
%Generating the load test using analysis of SUT data-flow and symbolic execution \cite{yang1996towards, zhang2011automatic} are examples of using source code analysis to generate load test and find performance-related issues.\\
Zhang \textit{et. al.} \cite{zhang2011automatic} present a symbolic execution-based approach using the source code for generating load tests.
Yang and Pollock \cite{yang1996towards} introduced a method for stress testing, limiting the stress test to parts of the modules that are more vulnerable to workloads. They used static analysis of the module's code to find these parts.

\paragraph{Modeling real usage}
%Extracting the usage pattern of real users and modeling their behavior based on form-oriented models \cite{draheim2006realistic, lutteroth2008modeling}, extracting workload characteristics and modeling the user behavior based on Extended Finite State Machines \cite{shams2006model} and Markov chains \cite{vogele2018wessbas} through monitoring submitted requests to SUT, and workload characterization through users clustering based on the business-level attributes extracted from usage data \cite{maddodi2018generating} are examples of the techniques used for modeling the realistic workload.\\
Draheim \textit{et. al.} \cite{draheim2006realistic} presents an approach for load testing of based on stochastic models of user behavior.
Lutteroth and Weber \cite{lutteroth2008modeling} provide a stochastic form-oriented load testing approach.
Shams \textit{et. al.} \cite{shams2006model} uses an application model-based approach that is an extension of Finite State Machines and models the user's behaviour.
V\"ogele \textit{et. al.} \cite{vogele2018wessbas} use Markov Chain for modeling user behaviour in workload generation.
All the named papers here proposed approaches for generating realistic workloads.
%\cite{maddodi2018generating} not written yet

\paragraph{Declarative methods}
% \cite{schulz2019behavior}: Declarative Performance Engineering aims at providing meth- ods and tools abstracting away the complexity of specifying and executing high quality performance evaluations, by providing ab- stractions and automation enabling performance engineering ac- tivity specification using declarative languages. The most promi- nent works in the area are by Walter et al. [22] and Ferme et al. [9].

%Using a declarative Domain Specific Language (DSL) to specify the performance testing process and a model-driven test execution framework \cite{ferme2018declarative, ferme2017towards, walter2016asking}, and also using a specific behavior-driven language,  to specify load testing process in combination with a declarative performance testing framework like BenchFlow \cite{schulz2019behavior, ferme2017towards} are examples of declarative techniques for performance and load testing.\\

Ferme and Pautasso \cite{ferme2018declarative} conduct performance tests using their model-driven framework that is programmed by a declarative domain-specific language (DSL) provided by them.
%we propose a declarative domain-specific language (DSL) for software performance testing and a model-driven framework that can be programmed using the mentioned language and drive the end-to-end process of executing performance tests.
Ferme and Pautasso \cite{ferme2017towards} also
use BenchFlow that is a declarative performance testing framework, to provide a tool for performance testing. This tool uses DSL for the test configuration.
%BenchFlow tool : Users can specify performance activities (e.g., standard performance tests) by relying on an expressive Domain Specific Language for objective-driven performance analysis.
Schulz \textit{et. al.} \cite{schulz2019behavior} generate load test using a declarative behavior-driven approach where load test specification is in natural language.
%\cite{walter2016asking} not written yet

\paragraph{Machine learning-assisted methods}
 %Machine learning techniques such as supervised and unsupervised algorithms are often intended to build models and knowledge patterns from the data, while in other techniques like reinforcement learning algorithms, the intelligent agent learns the way to accomplish an objective through interaction with the environment. 
 %Machine learning techniques have been frequently used for analyzing the resulted data from the load testing, using Bayesian Network to predict the reliability from the load testing data \cite{avritzer2008reliability}, anomaly detection based on analysis of metrics data, e.g., resource usage, using clustering techniques \cite{syer2011identifying}, identifying performance signature based on performance metrics data using supervised and unsupervised learning techniques \cite{malik2013automatic, malik2010automatic} are some examples of using machine learning techniques for analysis of load testing data.\\
 Some approaches in load testing context, use machine learning techniques for analyzing the data collected from load testing.
 For example, Malik \textit{et al.} \cite{malik2013automatic} use and compare supervised and unsupervised approaches for analyzing the load test data (resource utilization data) in order to detect performance deviation.
 %Avritzer \textit{et al.} \cite{avritzer2008reliability}\\
 Syer \textit{et al.} \cite{syer2011identifying} use the clustering method for detecting anomalies (threads with performance deviations) in the system based on the resource usage of the system.
 %Machine learning techniques have also been applied to the generation of performance test conditions in some studies. For example, RL together with symbolic execution has been applied to finding the worst-case execution path within a SUT in \cite{koo2019pyse},\\
 Koo \textit{et al.} \cite{koo2019pyse} provides a RL-based symbolic execution to detect worst-case execution paths in a program. Note that symbolic execution is mostly used in more computational programs manipulating integers and booleans.
 Grechanik \textit{et al.} \cite{grechanik2012automatically} presents a feedback-directed method for finding performance issues of a system by applying workloads on a SUT and analyzing the execution traces of the SUT to learn how to generate more efficient workloads.\\

\begin{table}[h!]
\caption{Overview of Related Work}
\begin{center}
\begin{tabular}{|p{2cm}|p{4cm}|p{6cm}|}
\hline
\textbf{Reference} & \textbf{Required Input}&\textbf{General Goal} \\
 \hline
 \cite{zhang2002automated,gu2009search,di2007search, garousi2010genetic, garousi2008traffic} & System model& Generate performance test cases using Petri nets, usage pattern model, and UML model\\
 \hline
 \cite{yang1996towards, zhang2011automatic} &Source Code & Finding performance requirements violation via static analysis and symbolic execution\\
 \hline
  \cite{draheim2006realistic, lutteroth2008modeling,shams2006model,vogele2018wessbas}&User behaviour model& User behaviour simulation-based load testing\\
 \hline
 \cite{ferme2018declarative, ferme2017towards, schulz2019behavior}& Instance Model of Domain-Specific Language & Propose Declarative methods for performance modeling and testing \\
 \hline
\cite{syer2011identifying, malik2013automatic} &Training set& Uses Machine learning-assisted methods for load test generation \\
 \hline
 \cite{koo2019pyse,grechanik2012automatically,ahmad2019exploratory}&System/program inputs& Finding worst-case performance issues using RL\\
 \hline
 \textit{This Thesis} & \textit{List of available transactions}&\textit{Generate optimal workloads that violates the performance requirements, using RL} \\
 \hline
\end{tabular}
\label{table: related_work}
\end{center}
\end{table}

%\textcolor{blue}{The paper "Exploratory Performance Testing Using Reinforcement Learning" \cite{ahmad2019exploratory}. I have to write down why is their paper bullshit.}\\

Ahmad \textit{et al.} \cite{ahmad2019exploratory} try to find the performance bottlenecks of the system using an RL approach named PerfXRL, which uses a DDQN algorithm. This is one of the more similar approaches to our approach recently published.
In their approach, each test scenario is a sequence of three constant requests to a web application. These requests have four variables in total, and the research aim is to find combinations of these four variables, which cause a performance violation. So the performance testing is done by executing test cases in which each test case is a sequence of three constant requests, and unlike our approach, no load testing is performed in this paper. They evaluate their approach by comparing the number of performance bottleneck request scenarios found by the PerfXRL approach with the number of performance bottleneck request scenarios found by a random approach. This comparison is made for different sizes of input value spaces. They show that for input value spaces bigger than a certain size (150000) the PerfXRL approach identifies more performance bottlenecks than the random approach\\

Unlike most of the mentioned approaches, our approach is model-free and does not require access to the source code or a system model for generating load tests. On the other hand, unlike many of the machine learning approaches, our proposed approach does not need previously collected data, and it learns to generate workload while interacting whit the system.
\\

\newpage
\section{Problem Formulation}\label{problem}

%\instructions{In this section you formulate and specify the three important things purpose, question and motivation. You should present the task in a clear way, both at a high level and in detail, and discuss why it is important. Explain assumptions and constraints. From the description of the task you can then formulate the purpose and the question. Keep in mind that when the purpose is met, the question should be answered. It is also important that purpose and motivation are linked. When the purpose and the question are clear, you can start developing the goals, the goals must be achieved to reach the goal. Each goal should be small, feasible and possible to evaluate.}
%\bigskip

%\textcolor{blue}{use chapter 8 from book \cite{berndtsson2007thesis} for writing this section.}

%\bigskip

The objective of this thesis is to propose and evaluate a load testing solution that is able to generate an efficient test workload, which results in meeting the intended objective of the testing, e.g., finding a target performance breaking point without access to system model or source code.
\subsection{Motivation and Problem}

With the increase of dependence on software in our daily lives, the correct functioning and efficiency of Enterprise Applications (EAs) delivering services over the internet are crucial to the industry. Software success not only depends on the correct functioning of the software system but is also dependent on how well are these functions performed i.e., non-functional properties like performance requirements). Performance bottlenecks can affect and harm performance requirements \cite{ibidunmoye2015performance, chandola2009anomaly}. Therefore, recognizing and repairing these bottlenecks are crucial.\\

The source of performance anomalies and bottlenecks can be application issues (i.e., source code, software updates, incorrect application conﬁguration), workload, the systems architecture and platforms, and system faults in systems resources and component (e.g. software bugs, environmental issues, and security violations.) \cite{ibidunmoye2015performance}. 
The source code would change during the continuous integration/delivery (CI/CD) process and software updates. The workload on the system is constantly changing, also the environmental issues and security conditions do not remain the same during the software's life cycle. Therefore the performance bottlenecks in the system will change during time, and it is not easy to follow the model-driven approaches for performance analysis. To perform performance analysis that can consider all mentioned causes of performance bottlenecks we can use model-free performance testing approaches.\\

In addition, an important activity in performance testing is the generation of suitable load scenarios to find the breaking point of the software under test. Manual workload generation approaches are heavily dependent on the tester’s experience and are highly prone to error. Such approaches for performance testing also consume substantial human resources and are dependent on many uncontrolled manual factors. The solution to this matter is using automated approaches. However, existing automated approaches for finding breaking points of the system heavily rely on the system's underlying performance model to generate load scenarios. In cases where the testers have no access to the underlying system models (describing the system), such approaches might not be applicable.\\

One other problem with existing automated approaches is that they do not reuse the data collected from previous load test generation for future similar cases, i.e. when the system should be tested again because of the changes made in the system during the time for maintenance, and scalability etc. There is a need for an automated, model-free approach for load scenario generation which can reuse learned policies and heuristics in similar cases.\\

%\todoB{\cite{zhang2011automatic}: Most of those efforts treat the program as a black box, focusing on increasing load by providing larger input sizes.  As we have shown, however, size is not all that matters.  The selection of the right combination of  input  values  can  deliver  an  equivalent  load  with  smaller  input  sizes  which can  reduce  testing  infrastructure  requirements,  can  provide  a  more  accurate  characterization  of scenarios where the system behaves poorly, can cover a range of resources, and may be helpful to identify anomalies that are exposed when traversing different execution paths. Yet, identifying such inputs can be extremely challenging since it requires an understanding of the program internals.  To address this challenge for smarter input selection, we developed }\\

Many model-free approaches for load generation, just keep increasing the load until performance issues appear in the system. The workload size is one factor that affects the performance, although the structure of the workload is another important factor. Selecting a certain combination of loads in the workload can lead to a violation of performance requirements and detecting performance anomalies with a smaller workload.
A well-structured smaller workload can more accurately detect the performance breaking points of the system with lower resources for simulating workloads. In addition, a well-structured smaller workload can result in increase coverage at the system-level.
Finding these specific workloads are difficult because it requires an understanding of the system's model.
\cite{zhang2011automatic}\\

%which can  reduce  testing  infrastructure  requirements,  can  provide  a  more  accurate  characterization  of scenarios where the system behaves poorly, can cover a range of resources, and may be helpful to identify anomalies that are exposed when traversing different execution paths. Yet, identifying such inputs can be extremely challenging since it requires an understanding of the program internals.  To address this challenge for smarter input selection, we developed \cite{zhang2011automatic}\\

Using model-free machine learning techniques such as model-free reinforcement learning \cite{sutton1998introduction} could be a solution to the problems mentioned above. In this approach, an intelligent agent can learn the optimal policy for performance analysis and load test scenarios that violate system performance. This method can be used independently of the system's and environment's state in different conditions, and it does not need to access the source code or system model. The learned policy could also be reused in further stages of the testing (e.g., regression testing).\\

%Our aim is to find workload performance bottlenecks of the system by generating workload scenarios that violate the performance requirements of the system. 

%We intend to generate workload-based test conditions for a system under test to reach a target error rate and response time threshold, without access to source code or system models, in an intelligent way i.e., based on model-free reinforcement learning methods. \\

\subsection{Research Goal and Questions}\label{sec:RQs}
%goal template:
%Analyze <Object(s) of study>
%for the purpose of <Purpose>
%with respect to their <Quality focus>
%from the point of view of the <Perspective>
%in the context of <Context>

%section 7.1 experiment in software engineering:
We intend to formulate a new method for load test generation using reinforcement learning and evaluate it by comparing it with random and baseline methods. Our technical contribution in this thesis is the formulation and development of an RL based agent, that will learn the optimal policy for load generation. We aim to evaluate the applicability and efficiency of our approach using an experiment research method. 
%Our contributions will add to the body of knowledge by reporting relevant quantitative results of the application of RL based approach for load testing.\\

%The object of study is the entity that is studied in the experiment. The object of study can be products, processes, resources, models, metrics or theories.
The \textit{object} of the study is an RL based load test scenario generation approach.
The \textit{purpose} is proposing and evaluating an automated, RL-based load test scenario generation tool. 
The \textit{quality focus} is the well-structured efficient test scenario, the final size of its workload, and the number of steps for generating the workload.
The \textit{perspective} is from the researcher’s and tester's point of view.
The experiment is run using an e-commerce website as a system under test.
Based on the GQM template for goal definition, presented by Basili and Rombach \cite{basili1988tame} our goal in this study is:\\

Formulate and analyze \textit{an RL-based load test approach}\

for the purpose of \textit{efficient\footnote{Efficient, in terms of optimal workload (workload size and number of steps for generating the workload).} load test generation}\

with respect to the  \textit{structure and size of the effective\footnote{Effective, in terms of causing the violation of performance requirements (error rate and response time thresholds).} workload, and the number of steps to generate it}\

from the point of view of a \textit{tester/researcher}\

in the context of \textit{an e-commerce website as a system under test}\\

%\noindent The objective of this thesis will be achieved by realizing the following main research goals:
%\begin{itemize}
    %\item \textit{\textbf{RG1:} To propose an automated model-free solution for load scenario generation}
   % \item \textit{\textbf{RG2:} To demonstrate the applicability of the proposed solution on a \todoB{experiment}}
%\end{itemize}

%\paragraph{RG1.}
%The first research goal is achieved by proposing and developing an Intelligent Load Runner as a reinforcement learning model-free approach shown in Figure \ref{ilr}.

%\textcolor{blue}{use chapter 8.5 for writing the tasks(objectives) from book \cite{berndtsson2007thesis}}

\noindent Based on our research goal we define the following research questions:\\

\begin{figure}[h]
\centering
\includegraphics[width=0.7\linewidth]{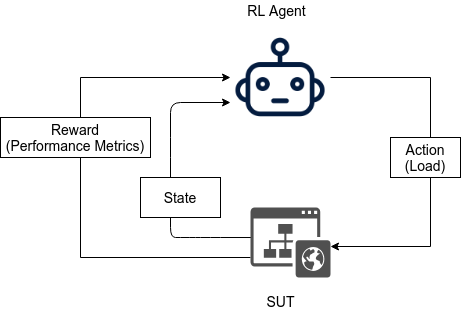}
\caption{Intelligent Load Runner}\label{ilr}
\end{figure}

\textbf{\textit{RQ1: How can the load test generation problem be formulated as an RL problem?}}
To solve the problem of load generation with reinforcement learning, a mapping should be done from the real-world problem to an RL problem environment and elements. The elements are the states, actions, and reward function (Figure \ref{ilr}). The aim of this research question is to find suitable definition of states, actions, and reward function in this problem.

\bigskip

\textbf{\textit{RQ2: Is the proposed RL-based approach\footnote{derived from RQ1} applicable for load generation?}} After formulating the problem into an RL context, it is essential to evaluate the applicability of the approach on a real-world SUT.
Answering this research question requires implementing the approach and setting up a SUT on which the generated load scenarios can be executed (see Section \ref{sec:sut}). 

\bigskip

\textbf{\textit{RQ3: What RL-based method is more efficient in the context of load generation?}}
Reinforcement learning can be applied using various algorithms like q-learning, SARSA (State-Action-Reward-State-Action), DQN, and Deep Deterministic Policy Gradient (DDPG). The aim of this research question is to choose at least two RL methods and find the most efficient (in terms of optimal) among them. In our case, we chose q-learning (a very basic RL algorithm) and DQN (an extended q-learning method). In addition, we also compare the results of the RL-based methods with a baseline and a random load generation methods.

\vspace{5mm}

%\textcolor{red}{I should be able to answer this question that for example why did you choose deep q-laerning among all other q-learning algorithms. I answerd this question in the modeling section.}

\bigskip

%performance metrics like error rate and response time

%putting the reinforcement learning algo
% using qlearning
% should I mention using dql?

%Talk about the proposed approach in detail here with figure and planned tasks

%\paragraph{\textcolor{blue}{Task3: choosing the EC algorithms I want to compare with}}

%\subparagraph{\textcolor{blue}{Task3: Finding a measure to compare our approach with a random approach: studying papers for this purpose}}

%\subparagraph{\textcolor{blue}{Task3: Implementing different RL approaches}}
%\subparagraph{\textcolor{blue}{Task4: Efficiency and sensitivity analysis}}
%\subparagraph{\textcolor{blue}{Task5: comparing different RL approaches}}
%\subparagraph{\textcolor{blue}{Task6: Implementing Evolutionary Computation Algs and comparing our results with their results}}

%\subparagraph{\textcolor{blue}{Task8: What are the existing load test generation approaches?}}
%\textcolor{blue}{In a project of this type, it is often necessary to implement the proposed solution, in order to demonstrate that it really does possess the proposed advantages. The goal of the implementation, then, is to demonstrate that the solution has certain properties, or that (under certain conditions) it behaves in a specific way. This implementation often needs to be compared with implementations of existing solutions, before conclusions can be drawn. The implementations of the existing solutions may or may not be done by yourself.}

%\subparagraph{\textcolor{blue}{Task8: Comparing the results of our solution to existing solutions}}
\newpage

\section{Methodology} 
\label{sec:method}

%\instructions{In this section, you will describe what scientific methods you have used and how you have approached the work itself. For each goal above, you identify a method for reaching the goal. The choice of method must be justified. For example, you may have done a mathematical model, used simulations, done an implementation that you have tested, or done an experiment that you might have evaluated using statistical methods. We intend here primarily that you describe the scientific methods you have used, but it is also good if you give a description of how you worked on the task. The Method section also answers why you did a certain way or why you used a certain tool. You should, therefore, not only describe "what" but also "why". Ask yourself the question: can the chosen method help me reach the set goals and thus answer the question?}

%\instructions{Choosing the right scientific method (s) is important for you to reach your goals, so this is a point that you should discuss with your supervisor at an early stage. Also, look in the literature for good descriptions of methods, and how to best write a Method section.}

%ab in meta text add something about the methodology that it is explained in section 5.2. Also, add that in order to evaluate the proposed models and achieve the research question  we used the case study research method explained 
\bigskip

%\textcolor{red}{I should do this section as Temirzhan Kozhakenov}\\

%ab in section 5.1 use present tense

A research method guides the research process in a step-by-step iterative manner. We use well-established research methods to realize our research goals. The core of our research method is the research process illustrated in Figure \ref{sec:rprocess}. The research process we used (to guide our research method) is a modification of the four steps research framework proposed by Holz \textit{et al.} \cite{holz2006research}. In the rest of this section, we presented our research process (in Section~\ref{sec:rprocess}) followed by a discussion on the research method used in Section~\ref{sec:rm}. Finally, we present the tools used for implementation in this thesis, in Section~\ref{sec:tools}.

\subsection{Research Process}\label{sec:rprocess}
In this subsection, we outline the research process that we are following throughout this thesis.

\begin{figure}[h]
\centering
\includegraphics[width=0.9\linewidth]{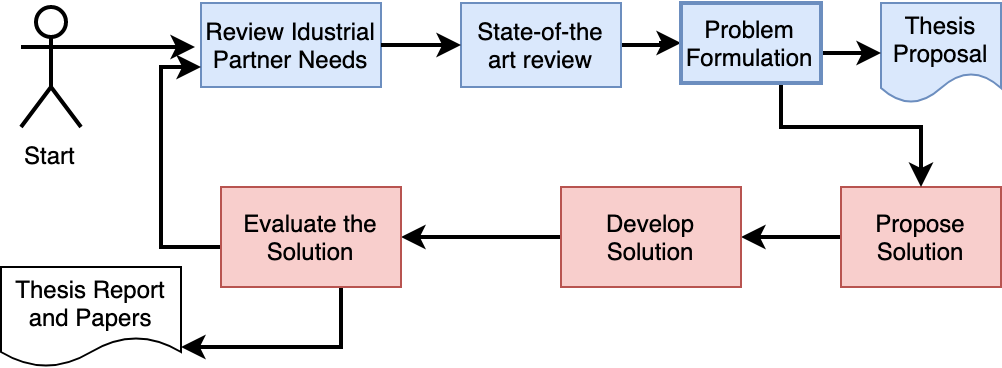}
\caption{Research Method}\label{rp}
\end{figure}

Our research process started with forming a suitable research goal and research questions (as formulated in Section~\ref{sec:RQs}). As discussed, the objective of our research is to propose and evaluate an automated model-free solution for load scenario generation.
The main objective and research goal were identified in collaboration with our industrial partner (RISE Research Institutes of Sweden AB) by reviewing their needs. We then identified specific challenges of the adoption of performance testing approaches with our industrial partner. We realized that existing approaches require knowledge of performance modeling and access to source code, which limits the adoption of such approaches.
We conducted a state-of-the-art review (some parts of it is presented in Section \ref{sec:background}) to identify the gaps in the literature. In the next step, we formulated an initial version of the problem which produced our thesis proposal.
In the next step of our research process, we formulated and initial RL based solution that does not require any underlying model of the system and can reuse the learned policy in the future. 
This formulated solution helped in realizing our primary research goal.
We then conducted an experiment to evaluate our solution on an e-commerce software system. Note that our research process was iterative and incremental.
%To address RQ1, we presented a mapping of the real problem to the RL environment (discussed in Section~\ref{sec:RQs}). Subsequently, to demonstrate the applicability of the RL-based approaches (RQ2), we developed and applied them on a SUT. Finally, to address RQ3 (to compare the effectiveness of two RL-based methods in the context of load generation), we executed and evaluated our proposed RL-approaches along with a baseline approach and a random approach for load test generation on a real-world experiment. To guide our research process and answer of research questions, we use the experiment research method following well-established guidelines for reporting~\cite{RuHo09}; discussion follows in the next section.

\subsection{Research Methodology}\label{sec:rm}
We conducted an experiment for answering our RQ3 following the guidelines presented by Wohlin \textit{et al.}~\cite{10.5555/2349018}. An experiment is a systematic formal research method in which the effects of all involved variables can be investigated in a controlled way. Thus, we can investigate the effect of our treatments (the different load test generation methods) on the outcome (size of workload generated which hit the thresholds, i.e., violates the performance requirements and the number of steps taken for generating this workload). 
Since our experiment's goal is to answer RQ3 (which requires quantitative data to answer), the experiment research method is helpful in obtaining quantitative data about the objectively measurable phenomenon.
In our case, the nature of the experiment is quantitative, i.e., comparing our RL-based load test generation approaches with a baseline and a random approach. The comparison is made based on the size of the workload generated that hits the defined error rate and response time thresholds. In addition, the comparison is also made based on the number of workload increment steps required for each approach to generate a workload that hits the thresholds.

\paragraph{Experiment Design}

The procedure of our experiment is explained in Section \ref{sec:procedure}.
Here we provide the standard definition of experiment terminologies in the guidelines~\cite{10.5555/2349018}, and we define them in our experiment:

\begin{itemize}
    \item Independent variables: \say{all variables in an experiment that are controlled and manipulated.}\cite{10.5555/2349018} In this experiment, the independent variables are the client machine generating workload, the client machine configuration, the network, the SUT server machine, and the SUT server configurations, and the parameters in Table \ref{table: load_tester_config_perparameters}.
    
    \item Dependent variables: \say{Those variables that we want to study to see the effect of the changes in the independent variables are called dependent variables.}\cite{10.5555/2349018} The dependent variables in this experiment are:
    \begin{itemize}
        \item size of the final workload generated that hits the defined error rate and response time thresholds. 
        %\item \hl{the structure of the final workload generated that hits the defined error rate and response time thresholds. }
        \item number of workload increment steps required to generate a workload that hits the thresholds.
    \end{itemize}
     
    \item Factors: one or more independent variables that the experiment studies the effect of changing them. The Factor, in our case, is the load test generation method.
    
    \item Treatment: \say{one particular value of a factor.}\cite{10.5555/2349018} The treatments in our experiment are a baseline method, a random method, a q-learning method, and a DQN method for our factor load test generation method.
  
    \item Subjects:  the subject, in our case, is the client machine generating workload. The properties of this machine are shown in Table \ref{table: machine_properties}.
    
    \item Objects: Instances that are used during the study. The object in our case is the SUT. The SUT is an e-commerce website explained in Section \ref{sec:website}.

\end{itemize}

%Case study a research methodology which explores a subject in depth. P.Runeson \textit{et al.}~\cite{RuHo09} define case study as: "an empirical method aimed at investigating contemporary phenomena in their context". some other related research methodologies are survey, experiment, and action research~\cite{RuHo09}.

%This case study has an exploratory and quantitative nature. Exploratory is one of the types in research classifications. 
%The purpose of an exploratory research is discovering what is happening in a context and expanding our perception and understanding of the subject.
%In our case, our research questions are of exploratory nature and an exploratory case study is a suitable choice of research method.
%Case studies are performed in real-world context which are more realistic, thus, in this methodology control over the phenomena decreases~\cite{RuHo09}.
%The methodology we used is a case study since we applied our load testing models on an open-source real-world SUT, on which we had low control on. However, we did had some control over how the SUT was deployed.

\subsection{Tools for the Implementation}\label{sec:tools}
Here we introduce the tools we used in our implementation and the reason for selecting them.

\paragraph{Apache Jmeter}
Apache JMeter is an open-source performance testing java application. It can test performance on static and dynamic resources. Apache JMeter can simulate heavy loads on a server, group of servers, network or object to test and measure performance metrics of the system under different load types. It is written in Java, and it allows us to use its libraries for executing our desired workloads in the implementation of our approach, which is written in Java. Additionally, JMeter has a simple and user-friendly GUI, which helps us easily generate JMX files containing the basic configurations needed for the workloads generated and executed in our load tester. 

\paragraph{WordPress and WooCommerce}
WordPress is a free and popular open-source content management system. It is written in PHP and paired with a MySQL database. We set up a website on WordPress as the SUT in the evaluation phase of our load testing approach. WordPress is very flexible and could be extended by using different plugins. WooComerce is an open-source e-commerce plugin for WordPress to create and manage online stores. We use WooComerce to turn the website into an e-commerce store

\paragraph{XAMPP}
XAMPP is one of the most common desktop servers. It is a lightweight Apache distribution for deploying local web servers for testing purposes. We create the WordPress website (SUT) using XAMPP.

\paragraph{RL4J}
In order to avoid possible implementation errors in implementing the DQN in one of our proposed approaches for load testing, we use an open-source library RL4J \cite{rl4j}. RL4J is a deep reinforcement learning library that is a part of the Deeplearning4j project \cite{Deeplearning4j} and released under an Apache 2.0 open-source license. Eclipse Deeplearning4j is a deep learning project written in Java and Scala. It is open-source, and it is integrated with Hadoop and Apache Spark and could be used on distributed GPUs and CPUs. Deeplearning4j is compatible with all java virtual machine language e.g., Scala, Clojure, or Kotlin. It includes deep neural network implementations with lots of parameters to be set by the users when training a network \cite{Deeplearning4j}. RL4j contains libraries for implementing DQN (Deep Q-learning with double DQN) and Async RL (A3C, Async NStepQlearning).

%\newpage
%\input{./outcome}
%\newpage
%\input{./etics}
\newpage
%\section{\instructions{``Description of the work'' }}

%\instructions{Following the sections above follows a description of what you have done. You should not use the heading above, but replace it with appropriate headings, depending on your work. The structure should be made clear through the section headings. A clear and clear logical structure and narrative flow are important. You should have advanced background knowledge that is necessary to understand how you solved the task, and define hypotheses and important concepts. The description of experiments should be such that the experiments can be repeated. If such a description becomes very long and detailed, you can put it in an appendix, see below.\\}

\section{Approach}
\label{sec:approach}

In this section, we propose our approach for intelligent load test generation using reinforcement learning methods. We answer RQ1 here and present the mapping of the real-world problem to an RL problem. We provide the details of our approach and the learning procedure for generating load test. 
In section \ref{sec:modeling_environment}, we provide the mapping of the optimal load generation problem to an RL problem, how we define the environment and the RL elements in the problem. Then in section \ref{sec:RL_method}, we present the RL methods that we use in our approach, which are q-learning and DQN, we also present the operating workflow for each method.

\subsection{Defining the environment and RL elements}\label{sec:modeling_environment}

In this section, we map the load test scenario elements to reinforcement learning elements and define the environment.

\paragraph{Agent and Environment}
As mentioned before, the goal of the agent is to attain the optimal policy, which is to find the most efficient workloads for testing the system's performance. For applying an RL-based approach to a problem, it is generally supposed that the environment is non-deterministic and also stationary upon transitions between the states of the system. The environment here is a server (the system under test) that
is unknown to the agent. The agent interacts with the SUT continuously, and the only information that the agent knows about the SUT is gained by the agent's observations from this interaction. The interactions are actions taken by the agent and the SUT's responses to these actions in the form of observations for the agent. In other works, the actions that our agent takes affects the SUT as the environment, and the SUT returns metrics to the agent, which affects the agent's next action.

\paragraph{States}
We define the states according to performance metrics. Error rate and response time are two performance metrics in load testing. These two are considered as the agent's observations of the environment. The two metrics define the agent's state; the average error rate and average response time returned from the environment (SUT) after the agent took the last action. The terminal states are the states with average response time or average error rate higher than a threshold. The average error rate range is 0 to \textit{error\_rate\_threshold} and the average response time rage is 0 to \textit{response\_time\_threshold} are divided into sections, each section determines one state.

\paragraph{Actions}
The action that the agent takes in each step is increasing the workload and applying it to the SUT (environment). The workload is generated based on the policy and the workload in the previous action. The workload contains several transactions in which each transaction has a specific workload, i.e., a specific number of threads executes each transaction. A transaction consists of multiple requests. A single thread represents a user (client) running the transaction and sending requests to the server (SUT). The action space is discrete, and the set of actions is the same for all the states. Each action increases the last workload applied to the SUT by increasing the workload of exactly one of the transactions. The workload of a transaction is increased by multiplying the previous workload in a constant ratio. The definition of actions is shown in equation \ref{eq:action} and equation \ref{eq:action2}:

\begin{equation} \label{eq:action}
Actions = \{\cup\ action_k,\ 1 \leq k \leq |\textit{Transactions}| \}
\end{equation}

\begin{equation} \label{eq:action2}
\begin{split}
action_k=\{\cup\ (W_t^{T_j})\ |\ & W_t^{T_j} = W_{n-1}^{T_j},\ \ if\  j\ne\ k,\\
           & W_t^{T_j}=  \alpha W_{n-1}^{T_j},\ if\ j=k,\\
           & T_j \in \textit{Transactions},\\
           & 1 \leq j \leq |\textit{Transactions}|\}
\end{split}
\end{equation}\\

\noindent Where $T_j$ indicates  transaction number $j$ among the set of transactions, $t$ is the current learning time step (iteration), $W_t^{T_j}$ is the workload of the transaction $T_j$ at time step  $t$, and $\alpha$ is the constant increasing ratio.

\paragraph{Reward Function}
The reward function takes an average error rate and average response time as input. The reward will increase as the average error rate and average response time increase. Consequently, the probability of the agent choosing actions which lead to a higher error rate and response time will increase. We define the reward function in equation \ref{eq:reward_function}.

\begin{equation}\label{eq:reward_function}
    R_t = (\frac{RT_t}{RT_{threshold}})^2+(\frac{ER_t}{ER_{threshold}})^2
\end{equation}\\

\noindent Where $R_t$ is the reward in time step $t$,  $RT_t$ is the average response time and $ER_t$ is the average error rate in time step $t$. And $RT_{threshold}$ and $ER_{threshold}$ indicate the response time and error rate threshold.

\subsection{Reinforcement Learning Method}\label{sec:RL_method}

In this section, we propose our RL solution to adaptive load test generation. We present our approach and explain the reinforcement learning algorithms that we chose for the approach, which are simple q-learning and DQN. We formulate the load test scenario in a reinforcement learning context and provide the architecture of our approach for each of the q-learning and DQN methods.\\

Algorithm \ref{Algorithm: Rl-driven load testing} shows a general overview of the RL method. We use two methods q-learning and DQN for the leaning phase in the Algorithm \ref{Algorithm: Rl-driven load testing} explained in sections \ref{qlearning_method} and \ref{DQN_method}.\\

\begin{algorithm}[H]
\SetAlgoLined
%\KwData{this text}
%\KwResult{how to write algorithm with \LaTeX2e }
\caption{Adaptive Reinforcement Learning-Driven load Testing}\label{Algorithm: Rl-driven load testing}
%\begin{algorithm2e}
%\begin{justifying}
\textbf{Required:} \(S, A, \alpha, \gamma\);\\
\textrm{Initialize q-values,\  \(Q(s,a)= 0\ \forall s \in \mathbb{S},\  \forall a \in \mathbb{A}\ \textrm{and}\ \epsilon=\upsilon\ ,0 < \upsilon <1\)};\\
%\textrm{Initial Learning};\\ 
%Initial Learning:\\
\While{ Not (initial convergence reached)}{
        Learning (with initial action selection strategy, e.g. $\epsilon$-greedy, initialized $\epsilon$)\;
    }

\textrm{Store the learned policy};\\
%\textrm{Start the transfer learning phase};\\
%\textbf{Transfer Learning:}\\
\textrm{Adapt the action selection strategy to transfer learning, i.e. tune parameter $\epsilon$ in $\epsilon$-greedy};\\
\While{true}{
Learning with adapted strategy (e.g., new value of $\epsilon$);\\
}
%\end{justifying}
%\end{algorithm2e}
\end{algorithm}

\subsubsection{Q-Learning}\label{qlearning_method}

As mentioned in section \ref{machine_learning}, q-learning is one of the basic reinforcement learning methods. 
Like other RL algorithms, q-learning seeks to find the policy which maximizes the total reward. 
The optimal policy here is extracted from the optimal q-function that is learned through the learning process by updating the q-table in each step. As mentioned before, q-tables store q-values, which get updated continuously. 
The q-value $q_\pi(s,a)$ of a state-action shows how good is to take action $a$ from state $s$. 
In each step, the agent is in a state and can perform one of the available actions from that state. In q-learning, the agent will take action with the maximum q-value among the available actions. 
As mentioned in section \ref{sec:background_RL}, choosing the action with the maximum q-value would satisfy the exploitation criteria. However, we also have to take random actions to satisfy the exploration criteria and be able to experience the actions with lower q-values, which have not been chosen before (therefore, their q-value is not updated and is low). 
Consequently, we use the decaying $\varepsilon$-greedy policy in which the $\varepsilon$ is big at the beginning of the learning and decays during the process. As mentioned before in Section~\ref{sec:background_RL}, $\varepsilon$ is a number in the range of 0 to 1. 
In each step, the probability of selecting the best actions is 1-$\varepsilon$, and a random action is selected by the probability of $\varepsilon$. 
After the action selection, we will observe the environment (in our case the SUT), and we will detect the next state and compute the reward then update the q-table with a new q-value for the previous state and the taken action. The q-learning algorithm is shown in Algorithm \ref{Algorithm: Q-learning)} \cite{sutton1998introduction}:\\ 

\begin{algorithm}[H]
\SetAlgoLined
\caption{Q-learning (off-policy TD control) for estimating $\pi \approx \pi_*$}
\label{Algorithm: Q-learning)}
\textrm{Algorithm parameter: step size $\alpha \in (0,1]$, small $\varepsilon > 0$}\\
\textrm{Initialise $Q(s,a)$, for all $s \in$ state space, $a \in A(s)$ arbitrarily except that \textit{Q(terminal,.)} = 0}\\

\For{each episode}{
    Initialize $S$\\
    \For{each step of episode}{
        Choose $A$ from $S$ using policy derived from Q (e.g., $\varepsilon$-greedy)\\
        Take action $A$, observe $R,S'$\\
    $Q(S,A) \xleftarrow{} Q(S,A)+ \alpha[R + \gamma \max\limits_{a}Q(S',a) -Q(S,A)]$\\        
    $S \xleftarrow{} S'$\\
    }
    until $S$ is terminal
}
\end{algorithm}

\bigskip

\begin{figure}[h]
\centering
\includegraphics[width=0.6\linewidth]{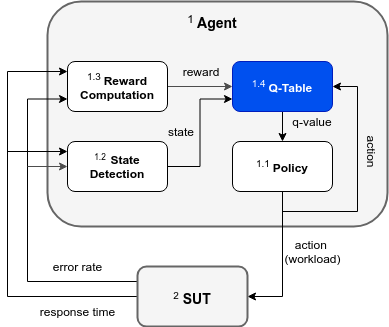}
\caption{The Q-learning approach architecture}\label{qlearning_model}
\end{figure}

\bigskip

\noindent Figure \ref{qlearning_model} illustrates the learning procedure in our approach:\\ 

\textit{\textbf{$^{1}$Agent.}} The purpose of the agent is to learn the optimum policy for generating load test scenarios that accomplish the objectives of load testing.  The agent has four components; Policy, State Detection, Reward Computation, and Q-Table.\\

\textit{\textbf{$^{1.1}$Policy.}} The policy which determines the next action is extracted from the Q-table based on the decaying $\varepsilon$-greedy approach; in each step, one action is selected among the available actions in the current state. As mentioned before each action is: increasing the workload of one of the transactions by a constant ratio, then applying the total workload of all transactions on the SUT concurrently.\\

\begin{figure}[h]
\centering
\includegraphics[width=0.5\linewidth, height=5cm]{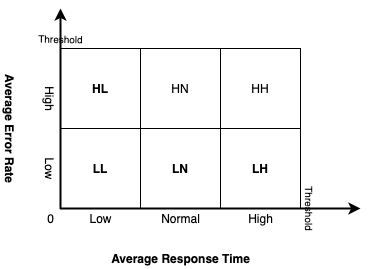}
\caption{States in the q-learning approach}\label{qlearning_states}
\end{figure}

\textit{\textbf{$^{1.2}$State Detection}} The state detection unit will detect the states based on the observations from the environment (i.e. SUT). The observations here are the error rate and response time. Each state is indicated by a range of average error rates and average response time. As Figure \ref{qlearning_states} shows, we define six states, each one covering a specific range in error rate and response time. We divided the [0, \textit{error\_rate\_threshold}] range into two sections and the [0, \textit{response\_time\_threshold}] range into three sections.\\

\textit{\textbf{$^{1.3}$Reward Computation}} The reward computation unit takes the error rate and response time as an input and calculates the reward based on them.\\

\textit{\textbf{$^{1.4}$Q-Table}} The q-table is where the q-values are stored. Each state-action has a q-value which will get updated by the gained reward after taking action from the state.\\

\textit{\textbf{$^{2}$SUT}} The environment in our case is the SUT in which the actions would apply to it, and it would react to the actions (i.e., applied workload). Then the agent receives observations from SUT, which are error rate and response time, and determine the state and reward based on them.

\subsubsection{Deep Q-Network}\label{DQN_method}

As mentioned in section \ref{machine_learning}, Deep Q-Network or DQN is an extension of q-learning. This method uses a function approximator instead of using a q-table. The function approximator, in this case, is a neural network. It approximates the q-values and refines this approximation (base on the rewards received each time after the agent taking action) instead of saving and retrieving the q-values from a q-table. Approximating q-values are beneficial when the state-action space is big. In this case, filling the q-table is not feasible and takes a long time. The benefit of using DQN is that it speeds up the learning process because 1) There is no need to store a big amount of data in the memory when the problem contains a large number of states and actions, 2) There is no need to learn the q-value of every single state-action and the learned q-values are generalized from the visited state-actions to the unvisited ones.\\

There are many function approximators (e.g., Linear combination of features, Neural Networks, Decision Tree, Nearest neighbor, Fourier/wavelet bases). Among the function approximators, neural networks are one of the function approximators which use gradient descent. Gradient descent is suitable for our data, which is not iid (Independent and Identically Distributed). The data is not iid because unlike supervised learning, in reinforcement learning values of the states near each other or the q-values of the state-action near each other are probably similar and the previous state is highly correlated with the previous state.\\

\begin{figure}[h]
\centering
\includegraphics[width=0.3\linewidth]{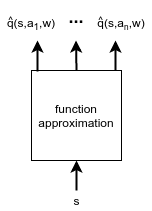}
\caption{DQN function approximation}\label{DQN_function_approximator}
\end{figure}

The DQN that we chose in our approach uses an ANN which takes a state as input and estimates the q-values of all the actions available from that state (Figure \ref{DQN_function_approximator}).\\ 

\begin{figure}[h]
\centering
\includegraphics[width=0.6\linewidth]{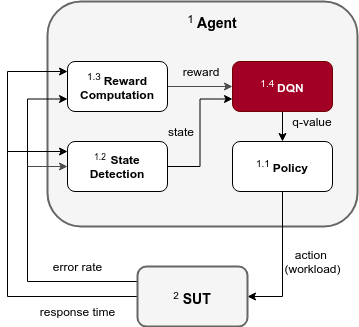}
\caption{The DQN approach architecture}\label{DQN_model}
\end{figure}

The architecture of our intelligent load runner approach with the DQN method is shown in Figure \ref{DQN_model}. The approach is the same as the q-learning approach except that it contains a \textit{$^{1.4}DQN$} unit instead of the \textit{Q-Table} unit. Also, in this approach, each state corresponds to a single response time, and error rate, and thus the number of states is equal to \textit{error\_rate\_threshold}$\times$\textit{response\_time\_threshold}. In each iteration, after receiving the reward, the DQN gets updated then the policy unit chooses an action based on the actions' q-value approximated by the DQN unit.\\

\newpage
\section{Evaluation}
\label{sec:evaluation}

%\todoB{The EVALUATION section of \cite{ahmad2019exploratory} exact text:\\
%In this section, we experimentally evaluate PerfXRL by comparing it against random testing. In random testing, we uniformly sample an input value combination without replacementt from the input space of the SUT for a given number of times. We choose random testing because it is a robust approach \cite{10.1145/1273463.1273476}, \cite{10.1109/TSE.1984.5010257} as compared to many other systematic testing approaches. Hamlet \cite{hamlet2006only} recommended using random testing for a large and irregular input space.
%In the rest of this section, we first describe the subject application used for evaluation, then we discuss the results.}

In this section, we explain the implementation setup of our proposed RL approaches for load test scenario generation, which answers RQ2. We explain the preparations for executing the implementation, i.e., the setup for the SUT, and also the procedure of how our experiment was setup. In addition, we evaluate our RL approaches by comparing them against a baseline and a random approach for generating the load tests in an experiment. 

\subsection{System Under Test Setup}\label{sec:sut}

Page loading time is an important factor in website's user experience, and page delays can result in big sale loss in e-commerce stores (online shops). Page loading time is dependent on performance requirements. Therefor performance requirements play a key role in e-commerce stores. We intend to use an open-source e-commerce store as a SUT to apply our proposed load testing method on it. Note that the e-commerce application is already being used in production by many users and in the real-world. Using an e-commerce store as the SUT makes it possible to send a variety of requests to the website as the workload. Requests like registering, logging in, visiting product pages, buying products online, etc. We cannot apply our approach to a running e-commerce store that provides real services to customers because load testing on the website will affect the website's performance and result in real sales loss. Therefore we will build our own e-commerce store.\\

In this section, we explain the implementation of the system under test in detail. We use XAMPP to deploy the SUT server and build a local e-commerce website using WordPress and WooCommerce.

\subsubsection{Server Setup}\label{sec:server_setup}

We deployed the SUT on a local server on a computer dedicated to this mean. We used a local server to avoid load testing through proxies. Otherwise, we will end up load testing the proxy server too, and the proxy may fail before the SUT server. Using a local server we can avoid possible effects of any in-between network equipment or server which may influence the test results.\\
%\todoB{++ We also set the SUT on a local server because network is the bottleneck in most cases: Which means it’s better to have medium machines with great network connectivity instead of powerful machines with low network capabilities.}\\

We deployed the SUT server using the XAMPP application on Ubuntu 16.04 operating system. As mentioned before XAMPP is a lightweight Apache distribution for deploying local web servers for testing purposes.\\

\begin{figure}[h]
\centering
\includegraphics[width=0.6\linewidth]{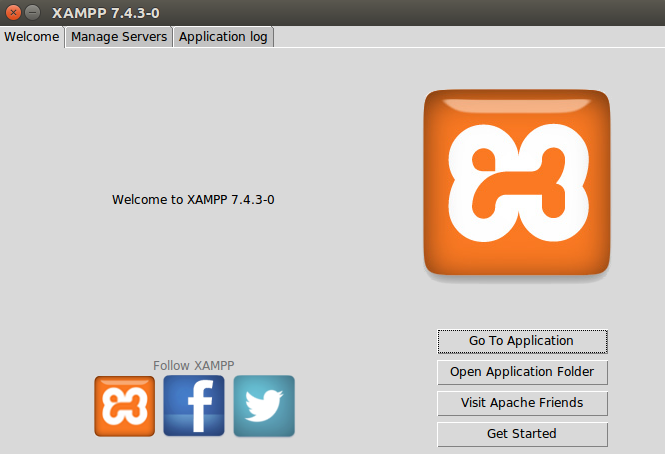}
\caption{XAMPP application}\label{xampp}
\end{figure}

To allocate our desired amount of resources to the SUT server, we use cproups. Cgroups or control groups, are a feature in  Linux kernel that makes it possible for the user to allocate resources e.g., CPU time, system memory, network bandwidth, or combinations of these resources among the collection of processes running on a system, and manage and put restrictions on the resources.

\begin{enumerate}
    
     \item We create a cgroup named "rlsutgroup" in the /etc/cgconfig.conf file:
        \smallskip
        \begin{lstlisting}[language=bash]
    group rlsutgroup {
        cpuset {
            cpuset.cpus = 0;
            cpuset.mems = 0;
        }
        memory{
            memory.limit_in_bytes = 2G;
        }
    }
\end{lstlisting}
\noindent We specify CPU number 0 and the memory node 0 to be accessed by the cgroup. We also set the maximum amount of user memory (including file cache) to 2 gigabytes (GB) for the cgroup.

% memory.memsw.limit_in_bytes = 2G didn't work in my laptop. Which sets the maximum amount for the sum of memory and swap usage.

%memory.soft_limit_in_bytes enables flexible sharing of memory. Under normal circumstances, control groups are allowed to use as much of the memory as needed, constrained only by their hard limits set with the memory.limit_in_bytes parameter. However, when the system detects memory contention or low memory, control groups are forced to restrict their consumption to their soft limits. To set the soft limit, for example, to 256 MB, execute: This parameter accepts the same suffixes as memory.limit_in_bytes to represent units. To have an effect, the soft limit must be set below the hard limit. If lowering the memory usage to the soft limit does not solve the contention, cgroups are pushed back as much as possible to make sure that one control group does not starve the others of memory. Note that soft limits take effect over a long time since they involve reclaiming memory for balancing between memory cgroups.

\bigskip

\item To move the SUT server process to the cgroup we created, we write the line below in /etc/cgrules.conf file:
\smallskip
\begin{lstlisting}[language=bash]
*:/opt/lampp/manager-linux-x64.run    cpuset,memory    rlsutgroup
\end{lstlisting}
 Where /opt/lampp/manager-linux-x64.run is the command for starting the Xampp server.
 \bigskip

\item To apply changes in cgconfig.conf and cgrules.conf we enter the commands below in ubuntu terminal:
\smallskip
\begin{lstlisting}[language=bash]
sudo cgconfigparser -l /etc/cgconfig.conf 
sudo cgrulesengd
\end{lstlisting}
\end{enumerate}

\subsubsection{Website Setup}\label{sec:website}

\begin{figure}[h]
\centering
\includegraphics[width=0.8\linewidth]{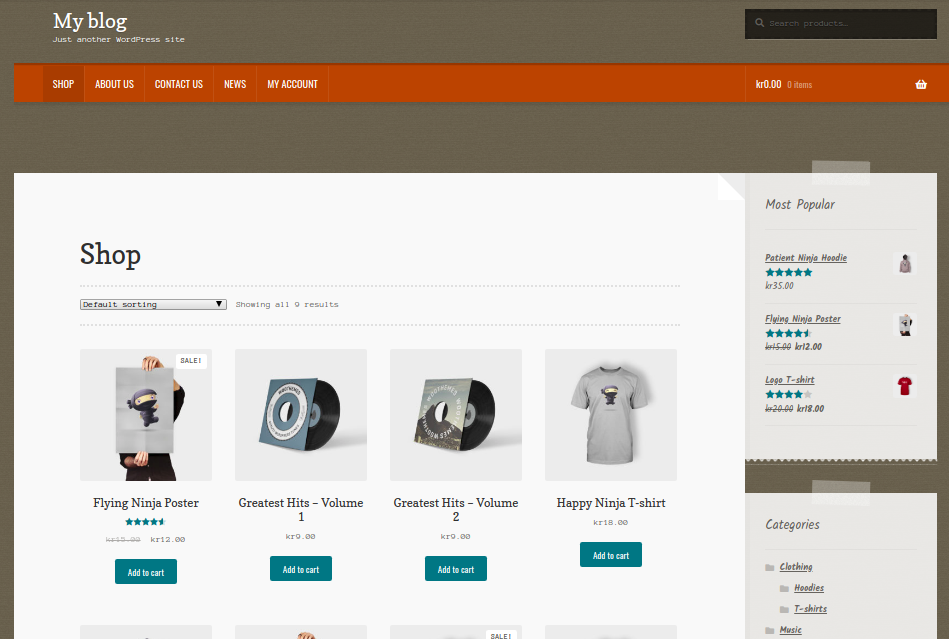}
\caption{SUT: E-commerce website}\label{sut}
\end{figure}

We set up an e-commerce website (Figure \ref{sut}) on WordPress using WooComerce. As mentioned before, WordPress is an open-source content management system, and WooComerce is an e-commerce plugin for WordPress to create and manage online stores and is being used by millions of users across the globe. A client can view products on the website, register or login to the website, add products to her cart, and checkout and order the products in her cart using PayPal or other options.

\subsection{Implementation}\label{sec:implementation}

In this section, we will first explain the structure of a workload and how they are generated using JMeter. Then we will provide the implementation details of our RL load tester.

\subsubsection{Workload Generation}

We use Apache JMeter as a load generation/execution tool in our implementation. As mentioned before, JMeter is a testing tool for generating and applying workload on servers. It generates the recommended workload by the tester agent, applies it on the SUT and, captures and measures the performance metrics of the SUT.\\

In each load test scenario, we execute a workload with eleven number of different transactions, in which each transaction has a specific size of workload executed by the JMeter threads. Each thread in JMeter indicates one user, and it is responsible for sending HTTP requests of one transaction to the SUT. While applying a workload on the SUT, for each transaction we will generate a number of JMeter threads equal to the size of that transactions specific workload (which is a variable we change during the workload generation process), and we will execute all the threads of every transaction in parallel in a specific ramp-up time.

\paragraph{Transactions}
We considered eleven different operations in the SUT shown in Table \ref{table: transactions} for generating workloads. A transaction is an operation and may have some functional prerequisite transactions. When a transaction is executed in the test, all of its prerequisite transactions would be executed sequentially in the specific order before the execution of the main transaction. Table \ref{table: prerequisite_transactions} shows the prerequisite transactions for each transaction. Each thread is responsible for executing a transaction and its prerequisite transactions sequentially. Nevertheless, all threads are executed in parallel.

\begin{table}[h!]
\caption{Common operations in an online shop}
\begin{center}
\begin{tabular}{|p{2.5 cm}|p{6cm}|}
 \hline
 \textbf{\textit{Operation}} & \textbf{\textit{Description}} \\
 \hline
 Home & Access to home page \\
 \hline
 Sign up page & Access to Sign up page\\
 \hline
 Sign up & Register and add a new user\\
 \hline
 Login page & Access to login page \\
 \hline
 Login & Sign in at the system\\
 \hline
 Search page & Access to search page\\
 \hline
 Select product & See the details of the selected product\\
 \hline
 Add to cart & Add the selected product to the cart\\
 \hline
 Payment & Access to payment page\\
 \hline
 Confirm & Confirm the order (payment) \\
 \hline
 Log out & Log out \\
 \hline
\end{tabular}
\label{table: transactions}
\end{center}
\end{table}

\begin{table}[h!]
\caption{Functions prerequisite transactions of each transaction}
\begin{center}
\begin{tabular}{|p{2.5 cm}|p{9cm}|}
 \hline
 \textbf{\textit{Transaction}} & \textbf{\textit{Prerequisite Transactions}} \\
 \hline
 Home & home page \\
 \hline
 Sign up page & home page $\rightarrow$ my account page\\
 \hline
 Sign up & home page $\rightarrow$ my account page $\rightarrow$ register\\
 \hline
 Login page & home page $\rightarrow$ my account page\\
 \hline
 Login & home page $\rightarrow$ my account page $\rightarrow$ login\\
 \hline
 Search page & home page\\
 \hline
 Select product & home page $\rightarrow$ select product\\
 \hline
 Add to cart &select product $\rightarrow$ add to cart\\
 \hline
 Payment & select product $\rightarrow$ add to cart $\rightarrow$ checkout\\
 \hline
 Confirm & select product $\rightarrow$ add to cart $\rightarrow$ checkout $\rightarrow$ PayPal page \\
 \hline
 Log out & my account page $\rightarrow$ logout \\
 \hline
\end{tabular}
\label{table: prerequisite_transactions}
\end{center}
\end{table}

\paragraph{JMeter Configuration}

We used apache-jmeter-5.2.1 for applying the workload on the SUT. JMeter could be run in a GUI mode or CLI (command line) mode.
JMeter Test Plans can also be created and executed through a java program.

A JMeter projects could be saved in a JMX file in the XML format. JMX or Java Management Extension is a standard framework for managing applications in java. It could be defined how to start, monitor, manage, and stop software components in a JMX file.

\subparagraph{Generating the Test Plan}
We use the GUI mode to generate a test; we setup JMeter to record user activities browsing the SUT. We do each of the transactions in table \ref{table: prerequisite_transactions} step by step and record the requests sent to the SUT using the JMeter.\\

Each test consists of some elements. All tests in JMeter should contain a Test Plan and Thread Groups. The steps for setting the elements are (Figure \ref{JMeter_test_plan}):

\begin{figure}[h]
\centering
\includegraphics[width=0.8\linewidth]{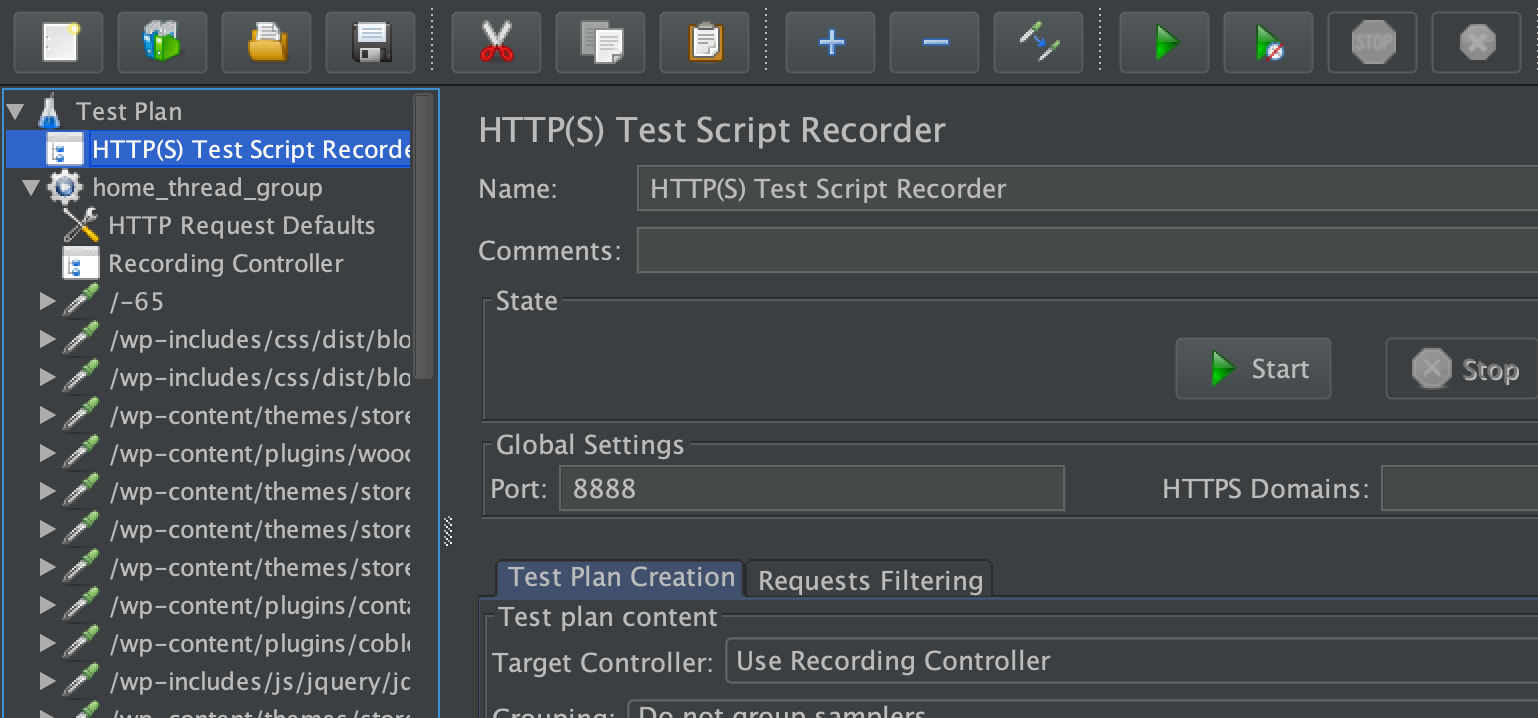}
\caption{JMeter Test Plan}\label{JMeter_test_plan}
\end{figure}

\begin{enumerate}
    \item Set the Test Plan element: In the Test Plan element, we check the option "Run tearDown Thread Groups after the shutdown of main threads" 
    \item Add Thread Group elements: In the Test Plan, we create a Thread Group element for one of the transactions:
    
    \smallskip
    \noindent right click on the Test Plan $\rightarrow$ Add $\rightarrow$ Threads (Users) $\rightarrow$ Thread Group
    
    \smallskip
    Each thread in a Thread Group simulates a user that sends requests to a server.
    
    \item Add HTTP Request Defaults element: HTTP Request Defaults is another JMeter element. The user can set default values for HTTP Request Samplers using HTTP Request Defaults. We add a HTTP Request Defaults to the Thread Group:
    
    \smallskip
    \noindent right click on the Thread Group $\rightarrow$ Add $\rightarrow$ Config Element $\rightarrow$ HTTP Request Defaults
    
    \smallskip
    \noindent We enter the website name under test in the "Server Name or IP" field in the HTTP Request Defaults control panel. In the "Timeout" box we set the "Connect" timeout to 30000 ms and the "Response" timeout to 120000 ms.

    \item Add Recording controller: JMeter can record the user activity and store it in a Recording controller. We add the Recording controller to the Thread Group:
    
    \smallskip
    \noindent right click on the Thread Group $\rightarrow$ Add $\rightarrow$ Logic Controller $\rightarrow$ Recording Controller
    
    \item Add HTTP(S) Test Script Recorder: HTTP(S) Test Script Recorder can record all the requests sent to a server. We add this element to the Test Plan:
    
    \smallskip
    \noindent right click on the Test Plan $\rightarrow$ Add $\rightarrow$ Non-Test Elements $\rightarrow$ HTTP(S) Test Script Recorder
    
    \smallskip
    \noindent We set the "Target Controller" field to "Test Plan $>$ Thread Group" where the recorded scripts will be added.
    
\end{enumerate}

After building the Test Plan;
\begin{enumerate}
    \item We change the proxy configuration of the browser and set the "HTTP Proxy" to "localhost" and the "Port" to the same port number in the HTTP(S) Test Script Recorder. \item Click the "Start" button in the HTTP(S) Test Script Recorder panel.
    \item Do the transaction step by step using the browser. 
    \item When the transaction is finished, we click the "Stop" button in the HTTP(S) Test Script Recorder panel.
    \item Save the JMeter project as a JMX file. 
\end{enumerate}

\begin{figure}[h]
\centering
\includegraphics[width=0.8\linewidth]{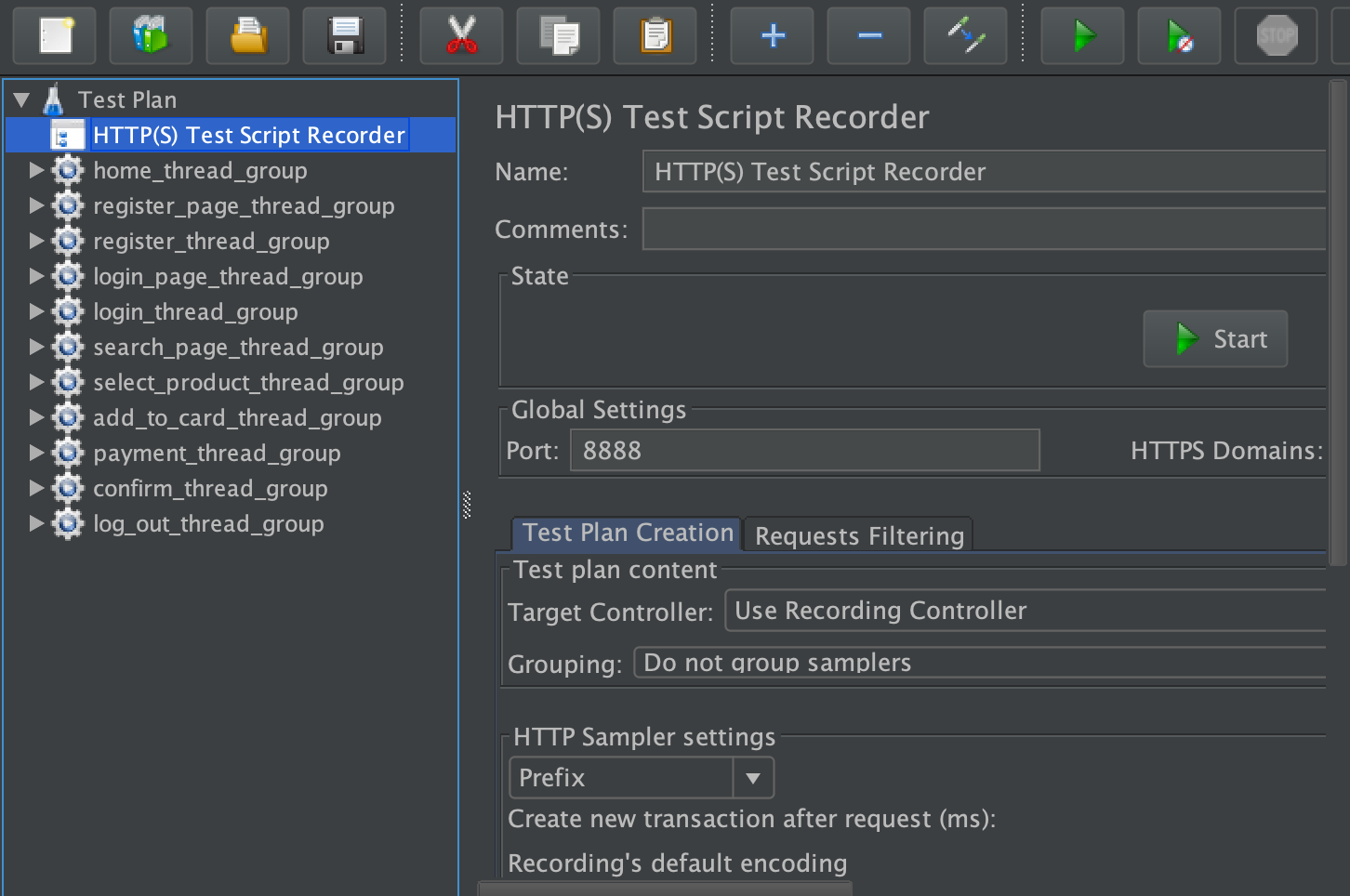}
\caption{JMeter Thread Groups}\label{JMeter_thread_groups}
\end{figure}

We do all the steps above for each transaction in table \ref{table: prerequisite_transactions}, and in the end, we integrate all of the thread groups in a single JMX file (Figure \ref{JMeter_thread_groups}) to be used for applying the workload by the agent. When executing the final JMX file, all the Thread Groups will start the same time and will execute concurrently.

\subparagraph{Executing the Test Plan}
We do not use JMeter GUI or CLI for executing the generated Test Plan. Instead, the Test Plan is executed from java code. The java program is the implementation of the RL load test generation wherein each step of the agent's learning process, the Test Plan is executed.\\

To run the JMeter Test Plan, first we increase the JMeter heap size to be able to generate larger workloads described as below:
\begin{itemize}
    \item Go to the apache-jmeter-5.2.1/bin directory
    \item  Open JMeter startup script
    \item Find the line HEAP="-Xms1g -Xmx1g"
    \item Change the maximum value to -Xmx4g
\end{itemize}

We load the JMX file in the java program by importing Apache JMeter packages:

\begin{lstlisting}[language=Java]
testPlanTree = SaveService.loadTree(new File (myTest.jmx"));
\end{lstlisting}

\noindent And reset the Thread Group parameters for each step of applying the workload after an action is taken by the agent ( i.e., one of the transactions workload is increased). There are three parameters to set for each Thread Group:

\begin{itemize}
    \item number of threads: The number of threads (workload) in the Thread Group. 
    \item number of loops: The number of times that the Thread Group is executed.
    \item ramp-up time: the time that takes for all threads of the Thread Group to get up and running.
\end{itemize}
\smallskip

%\todoB{did I set the time limitation for the last test plan?yes I did}\\

\noindent For each Thread Group, we set the number of threads equal to the workload of that transaction in the RL agent, and the ramp-up time equal to the workload divided by a ratio "threadPerSecond" (which we set it equal to 10). The number of loops is set to 1 for all Thread Groups.

\begin{lstlisting}[language=Java]
threadGroup.setNumThreads(transaction[i].workLoad);
threadGroup.setRampUp(transaction[i].workLoad/threadPerSecond);
((LoopController) threadGroup.getSamplerController()).setLoops(1);
\end{lstlisting}

\noindent Then we run the Test Plan:
\begin{lstlisting}[language=Java]
jmeter.run();
\end{lstlisting}

%%\paragraph{\todoB{System attributes and the maximum number of virtual threads I can run on the system}}
%\todoB{So do we need distributed testing on multiple systems?}
%\todoB{https://www.blazemeter.com/blog/whats-the-max-number-of-users-you-can-test-on-jmeter/: what’s the maximum number of users you can test on Apache JMeter™? Limits usually depend on many factors, and the same goes for JMeter limitations. JMeter limits depend on the capabilities of your machine and network, the complexity of your performance scripts, the targeted number of simulated users and so forth.}
%\todoB{
%\begin{itemize}
 %   \item Tip 1: Use JMeter Listeners for Debugging Purposes Only: JMeter listeners generate a lot of redundant load on your local machine, and they create lots of objects in the heap that might occupy most of the heap space. There are many other ways to monitor your scripts without generating a heavy load on your local machine. In this article you can find three ways to do it right. 
  %  \item Tip 2: Run JMeter Tests in Non-GUI Mode
   % \item Tip 3: Increase JMeter Heap Space to Generate a Larger Load
    %\item Tip 4: Use LAN instead of Wi-Fi Connections When Running Load Tests
%\end{itemize}
%}

To be able to apply a large number of concurrent requests, we should execute the program on a device with high memory/CPU resources. Table \ref{table: machine_properties} shows the properties of the device we used.

\begin{table}[h!]
\caption{Hardware Overview of the Machine used for executing Load Generation Client}
\begin{center}
\begin{tabular}{|p{4.5 cm}|p{4cm}|}
 \hline
 Model Name & MacBook Pro\\
 \hline
  Model Identifier & MacBookPro12,1\\
  \hline
 Processor Name & Dual-Core Intel Core i7\\
 \hline
  Processor Speed &    3,1 GHz\\
 \hline
  Number of Processors & 1\\
 \hline
  Total Number of Cores & 2\\
 \hline
  L2 Cache (per Core) &    256 KB\\
 \hline
  L3 Cache & 4 MB\\
 \hline
  Hyper-Threading Technology & Enabled\\
 \hline
  Memory & 16 GB\\
 \hline
\end{tabular}
\label{table: machine_properties}
\end{center}
\end{table}

\subsubsection{Q-Learning Implementation}

\begin{figure}[h]
\centering
\includegraphics[width=0.7\linewidth]{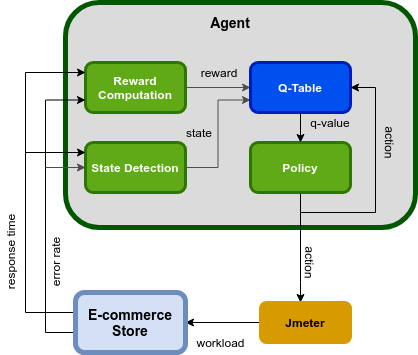}
\caption{Q-learning implementation architecture of Intelligent Load Runner}\label{qlearning_implimentation_architecture}
\end{figure}

We implemented the agent in a java program based on the approach explained in Section \ref{qlearning_method}. As shown in Figure \ref{qlearning_implimentation_architecture}, we have a module for each of Q-Table, Policy (action selection), Reward Computation, and State Detection. How the modules communicate with each other is explained in Section \ref{sec:procedure}.\\

\subsubsection{DQN Implementation}

We use the library RL4J \cite{rl4j} in our DQN approach implementation. As mentioned before, RL4J is a deep reinforcement learning library that contains libraries for implementing DQN (Deep Q-learning with double DQN).

\paragraph{Prerequirements for deeplearning4j}
%\textcolor{blue}{https://deeplearning4j.org/docs/latest/deeplearning4j-quickstart: \\}

Prerequisites:
\begin{itemize}
    \item Java: JDK 1.7 or later should be installed. (Only 64-Bit versions are supported)
    \item Apache Maven: Maven is a dependency manager for Java application.
    \item IntelliJ or Eclipse: IntelliJ and Eclipse are Integrated Development Environments (IDE) that help to work with deeplearning4j and configuring neural networks easier. IntelliJ is recommended for using the deeplearning4j library.
    \item Git: to clone deeplearning4j examples.
\end{itemize}

\paragraph{Configuring and training a DQN agent using RL4J}

\begin{enumerate}

\item Create an action space for the mission:
\begin{lstlisting}[language=Java]
    DiscreteSpace actionSpace = new DiscreteSpace(numberOfTransactions);
\end{lstlisting}

\item Create an observation space for the mission:
\begin{lstlisting}[language=Java]
    SUTObservationSpace observationSpace = 
        new SUTObservationSpace(maxResposeTimeThreshold,maxErrorRateThreshold);
\end{lstlisting}

\item Create an MDP wrapper:
\begin{lstlisting}[language=Java]
    ILRMDP mdp = new ILRMDP(maxResponseTimeThreshold, maxErrorRateThreshold
    , csvWriter);
\end{lstlisting}

\item Create a DQN:
\begin{lstlisting}[language=Java]
    public static DQNFactoryStdDense.Configuration LOAD_TEST_NET =
        DQNFactoryStdDense.Configuration.builder().l2(0.01).updater(
        new Adam(learningRate)).numLayer(3).numHiddenNodes(16).build();
\end{lstlisting}

\item Create a Q-learning configuration by specifying hyperparameters:
\begin{lstlisting}[language=Java]
    public static QLearning.QLConfiguration LOAD_TEST_QL =
        new QLearning.QLConfiguration(
            seed,
            maxEpochStep,
            maxStep,
            expRepMaxSize,
            batchSize,
            targetDqnUpdateFreq,
            updateStart,
            rewardFactor,
            gamma,
            errorClamp,
            minEpsilon,
            epsilonNbStep,
            doubleDQN
        );
\end{lstlisting}

\item Create the DQN:
\begin{lstlisting}[language=Java]
    Learning<QualityMeasures, Integer, DiscreteSpace, IDQN> dql = 
        new QLearningDiscreteDense<QualityMeasures>(mdp, LOAD_TEST_NET
        , LOAD_TEST_QL, manager);

\end{lstlisting}

\item{Train the DQN}
\begin{lstlisting}[language=Java]
dql.train();
\end{lstlisting}

\end{enumerate}

\subparagraph{Q-learning hyperparameters of the DQN}\label{hyperparameters}
The Q-learning configuration hyperparameters are \cite{raj2019java,ReinfLearnPlayPixels}:

\begin{itemize}
    \item \verb|maxEpochStep|: Each epoch is equivalent to an episode in the learning algorithm. \verb|maxEpochStep| is 
    the maximum number of steps allowed in each episode (epoch).
    \item \verb|maxStep|: The maximum number of total iterations (the summation of steps in all episodes) in the learning. Training will finish when the number of steps exceeds \verb|maxStep|.
    \item \verb|expRepMaxSize|: The maximum size of experience replay. The number of past transitions that the agent will take the next action based on them is experience replay. Experience replay is explained in detail in section \ref{sec:background}.
   
    \item \verb|batchSize|: The number of steps which the neural network would update its weights after executing them.\
    %\todoB{write down how did you choose the batch size based on these url:}\url{https://machinelearningmastery.com/gentle-introduction-mini-batch-gradient-descent-configure-batch-size/} and \url{https://machinelearningmastery.com/how-to-control-the-speed-and-stability-of-training-neural-networks-with-gradient-descent-batch-size/}\\
    
    We choose the batch size equal to 1 because each sample in RL is dependent on the previous sample, so the network should be updated per sample (in our case, each learning step).

    \item \verb|targetDqnUpdateFreq|: In double DQN the target network is frozen for \verb|targetDqnUpdateFreq| number of steps and it would update after \verb|targetDqnUpdateFreq| steps from the online network. The state-action values are computed (the evaluation) based on the target network to stabalize the learning.
    \item \verb|updateStart|: The number of no-operation (do nothing) moves before starting the learning to make the learning start with a random configuration. The agent will conduct the same sequence of actions at the beginning of each episode instead of learning to take the next action based on the current state if it starts with the same configuration each time.
    % If the agent starts the game, in the same way, every time, then the agent will memorize the sequence of actions rather than learning to take the next action based on the current state.
    \item \verb|rewardFactor|: Reward factor is an important hyperparameter that should be considered carefully. It significantly affects the efficiency of learning. This factor scales the rewards, so the Q-values will be lower (if the range is [-1; 1] it is similar to normalization).
    \item \verb|gamma|: The discount factor.
    \item \verb|errorClamp|: This parameter will clip (bound between two limit values) the loss function (TD-error) based on the output in the backpropagation. For example if \verb|errorClamp|=1, then the gradient is bounded to the range (-1,1).
    \item \verb|minEpsilon|: Epsilon is the derivative of the loss function with respect to the activation function's output. The epsilon is used to compute the gradients for every activation node in backpropagation.
    
    %isnt this epsilon for the epsilon greedy purpose? and if it is for the bakcpropagation, it is the ratio of the derivative not the actual derivative\\
    
    \item \verb|epsilonNbStep|: After \verb|epsilonNbStep| number of steps, the epsilon will be decreased to \verb|minEpsilon|.
    \item \verb|doubleDQN|: This value should be set to \textit{True} to enable double DQN.
\end{itemize}

\noindent We set the values of the hyperparameters as shown in Table \ref{table: hyperparameters}.

\begin{figure}[h]
\centering
\includegraphics[width=0.7\linewidth, height=9cm]{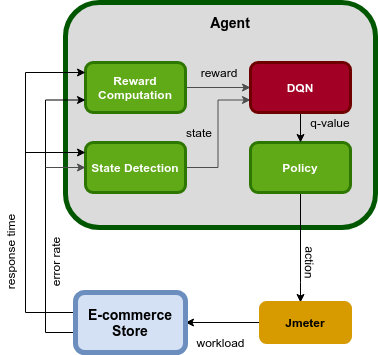}
\caption{DQN implementation architecture of Intelligent Load Runner}\label{DQN_implimentation_architecture}
\end{figure}

\subsection{Experiment Procedure}\label{sec:procedure}

We separately executed our RL approaches for load generation (shown in Figure \ref{qlearning_implimentation_architecture} and Figure \ref{DQN_implimentation_architecture}) on the SUT. 
We also executed a baseline load generation approach and a random load generation approach on the SUT to evaluate the efficiency of the proposed RL approach against them. 
All approaches were executed in several episodes, each episode consisting of several steps.
In the baseline approach, in each step, the workload size of all transactions was incremented.
We chose a baseline approach to compare the final size of the generated workload (that hit the error rate or response time thresholds) of the baseline approach with our proposed RL approaches. 
In the random approach, in each step, a random transaction was chosen, and the size of its workload was increased (unlike the RL approaches where the transaction was selected based on the policy).
The reason behind choosing a random approach was that random testing is found robust \cite{10.1145/1273463.1273476}, \cite{10.1109/TSE.1984.5010257} among many other systematic testing approaches and is a good criterion.\\

\begin{figure}[h]
\centering
\includegraphics[width=0.6\linewidth]{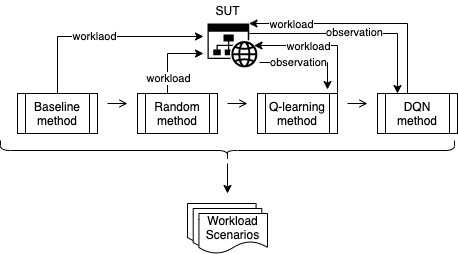}
\caption{Procedure of executing the methods}\label{procedure}
\end{figure}

The SUT was deployed on a local server, an ASUS K46 computer with a Ubuntu 16.04 operating system, dedicated 1 CPU, and 2 GB memory to the SUT (as mentioned in section \ref{sec:server_setup}).
During the execution of the methods, the system was logging necessary data for the evaluation metrics. 
An overview of the procedure is shown in Figure~\ref{procedure}.
We further explain the evaluation metrics used and the procedure for executing each approach. 

\paragraph{Evaluation metrics}
The evaluation metrics are the average error rate, average response time, size of the final effective workload, and number of steps for generating the effective workload. We can not conclude from a quick response time that the SUT is operating fine and fast, and we should consider the error rate too. Since the servers are quick at delivering error pages, we may get low response times with high error rates in some situations. Consequently, not only we put a threshold on the response time, but also we put a threshold on the error rate.

\paragraph{General configuration}

We execute each approach for about 40 episodes. Each episode consisted of several steps. 
In each step, a workload was generated and applied to the SUT. The workload, response time, and error rate were logged for each step.
The episodes continued until the observed error rate and response time hit the threshold. Every two continuous episodes were executed with a 5-minute delay between them, which allowed the server to go back to its normal state. Table \ref{table: load_tester_config_perparameters} shows the configuration values used for all the approaches.

\begin{table}[h!]
\caption{Load Tester Configuration Values}
\begin{center}
\begin{tabular}{|p{7.5 cm}|p{1.5cm}|}
 \hline
 \textbf{Parameter} & \textbf{Value}\\
 \hline
 average response time threshold & 1500ms\\
 \hline
 average error rate threshold & 0.2\\
 \hline
 delay between executing two continuous episodes & 5\\
 \hline
 number of started threads per second & 10\\
 \hline
 initial workload per transaction & 3\\
 \hline
 transaction workload increasing step ration & 1/3\\
 \hline
\end{tabular}
\label{table: load_tester_config_perparameters}
\end{center}
\end{table}

\paragraph{Baseline approach}
We executed the baseline approach for 40 episodes.
In each step of an episode in this approach:

\begin{enumerate}
    \item The workload of all transitions were increased by $1/3$ of their current workload.
    \item The Test Plan was loaded from the JMX file, and after setting the new values for each transaction workload, the Test Plan was executed, and the workload was applied on the SUT.
\end{enumerate}

\paragraph{Random approach} 
We executed the random approach for 40 episodes.
In each step of an episode in this approach:

\begin{enumerate}
    \item A transition was chosen randomly, and its workload was increased by $1/3$ its current workload.
    \item The Test Plan was loaded from the JMX file, and after setting the new values for each transaction workload, the Test Plan was executed, and the workload was applied on the SUT.
\end{enumerate}

\paragraph{Q-learning and DQN approaches}

We executed the q-learning approach for 40 episodes. However, the DQN approach was executed for 47 episodes. The reason behind this is that the number of episodes was not configurable in the DQN implemented using the RL4J library, and instead, the number of steps was a configurable parameter. Therefore we configured the number of steps equal to a value (shown in table \ref{table: hyperparameters}) that we approximated to be executed in around 40 episodes.
In each episode, the agent started from an initial state, which was detected by applying an initial workload on the SUT and observing the average error rate and response time. The initial q-values in the q-table/q-network were set to 0. Each episode consisted of several learning steps. In each learning step:

\begin{enumerate}
    \item An action was chosen according to the policy; one of the transactions workload would increase by $1/3$ its current workload.
    \item The Test Plan was loaded from the JMX file, and after setting the new values for each transaction workload, the Test Plan was executed, and the workload was applied on the SUT.
    \item Based on the observations (average error rate and average response time), the new reward and new state were detected, and the q-table or q-network got updated.
\end{enumerate}

%\noindent The execution delay between every two continues episodes allowed the server to go back to its initial state.\\

In the q-learning approach, we set the learning rate and discount factor both equal to 0.5, which are the q-learning attributes.
In the DQN approach, we set the values of the hyperparameters, as shown in Table \ref{table: hyperparameters}.

\begin{table}[h!]
\caption{Hyperparameters Configuration for DQN}
\begin{center}
\begin{tabular}{|p{4 cm}|p{1.5cm}|}
 \hline
 \textbf{Hyperparameter} & \textbf{Value}\\
 \hline
 maxEpochStep & 30\\
 \hline
 maxStep & 450\\
 \hline
 expRepMaxSize & 450\\
 \hline
 batchSize & 1\\
 \hline
 targetDqnUpdateFreq & 10\\
 \hline
 updateStart & 1\\
 \hline
 rewardFactor & 0.1\\
 \hline
 gamma & 0.5\\
 \hline
 errorClamp & 10.0\\
 \hline
 minEpsilon & 0.1f\\
 \hline
 epsilonNbStep & 400\\
 \hline
 doubleDQN & true\\
 \hline
\end{tabular}
\label{table: hyperparameters}
\end{center}
\end{table}

\newpage
\section{Results}
\label{sec:results}

%\instructions{Here you can, for example, present results of experiments, evidence, analysis of data etc. Your results must be described so clearly that a reader can judge them. You should also explain and analyze the results.}
%
This section presents the results of the experiment conducted to evaluate the efficiency of the baseline, random, q-learning, and DQN approaches. This section is focused on answering  RQ3. %As discussed, we also included a baseline approach and a random approach in our experiment, and this section also includes the results from those baselines.

\paragraph{Results of the Baseline Approach} 

\begin{figure}[h]
\label{baseline}
\centering
\subfloat[Number of Steps per Episode]{
    \label{steps_baseline}
    \framebox{\includegraphics[width=0.48\linewidth]{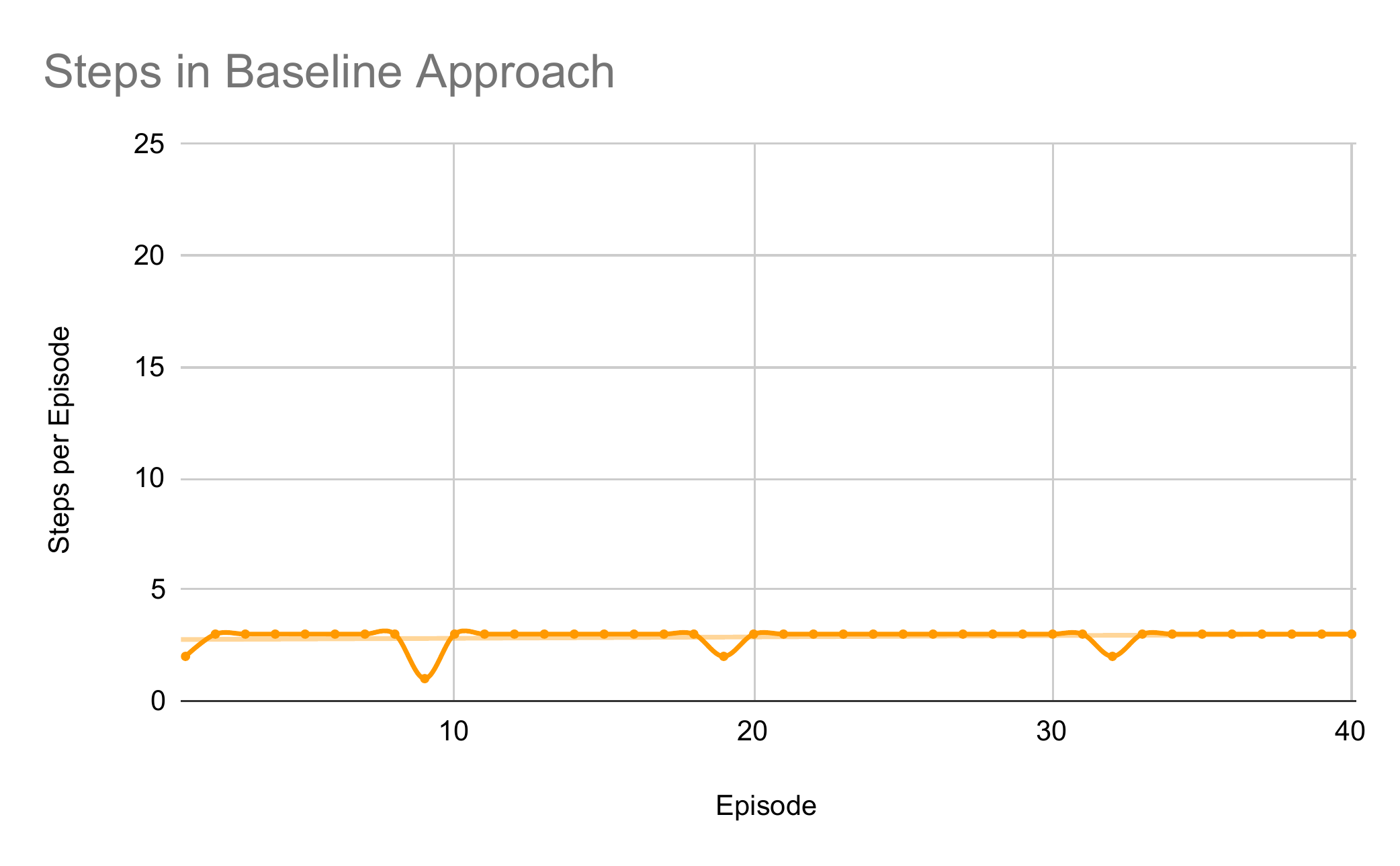}} } 
\subfloat[Final Workload Size per Episode]{
    \label{final_workload_baseline}
    \framebox{\includegraphics[width=0.48\linewidth]{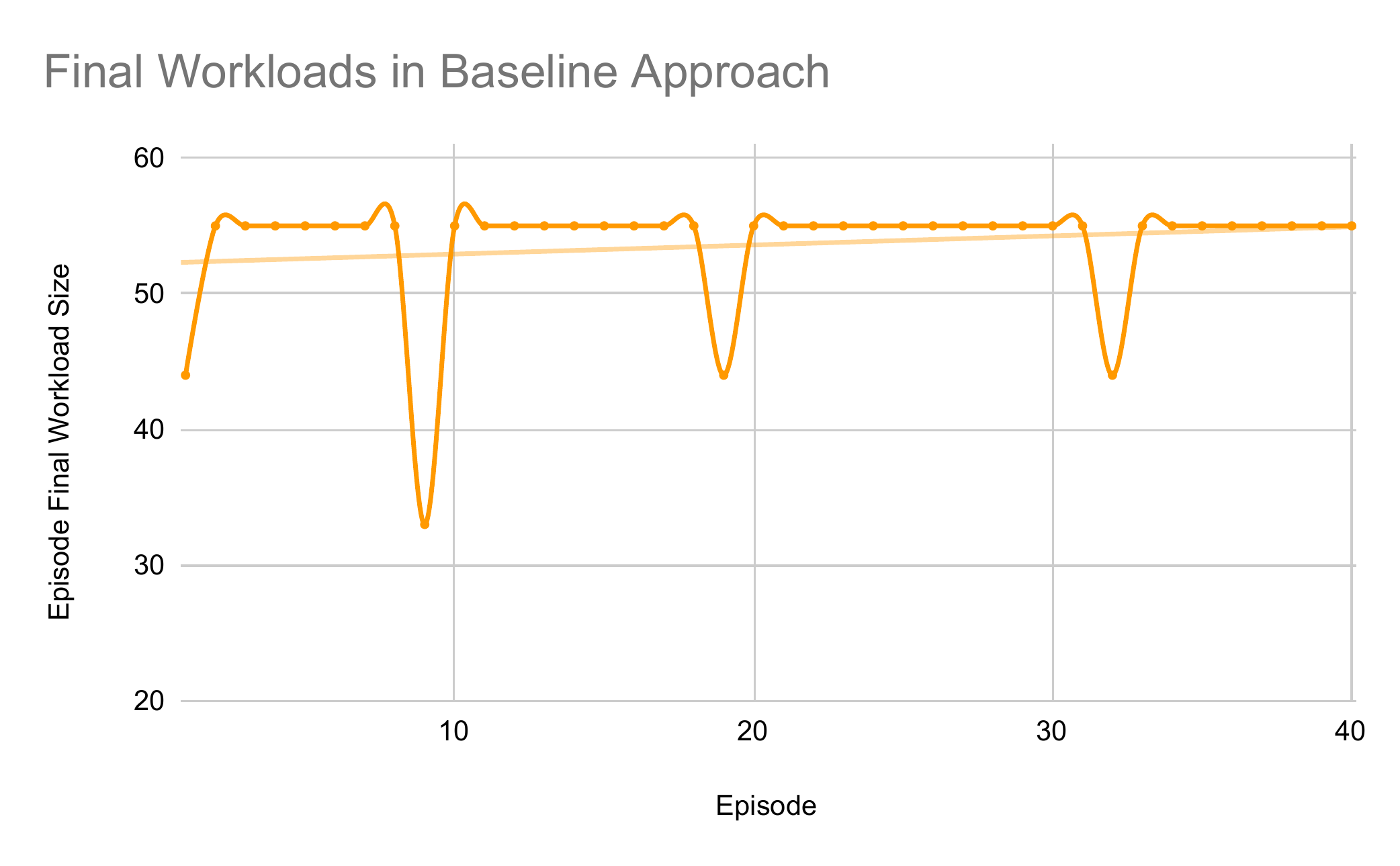}}}
\caption{Baseline Approach}
\end{figure}

The baseline approach was executed for 40 episodes.
In Figure~\ref{steps_baseline}, the episodes are plotted on X-Axis, and the Y-Axis shows the number of steps in each episode that is needed for generating the workload that hit the response time or error rate threshold. As it can be seen from Figure~\ref{steps_baseline}, the trend line for the baseline approach stays between zero to five.
It means that the baseline approach took fewer steps in generating the final workload that hits the thresholds. However, the increment of the workload size in each step was very high because it was applied to all transactions (as shown in Figure~\ref{final_workload_baseline}).% That is why the final workload is generated after a few big incremental steps.\\

In Figure~\ref{final_workload_baseline}, the episodes of baseline approach are shown on the X-Axis, and the size of the final workload that hits the response time or error rate threshold is shown on the Y-Axis.
As it can be seen in Figure~\ref{final_workload_baseline}, the trendline for the baseline approach consistently stays between 50 to 60. This means that in the majority of the episodes, the baseline approach was mostly able to hit the threshold with bigger workload sizes.

\paragraph{Results of the Random Approach} 

\begin{figure}[h]
\label{random}
\centering
\subfloat[Number of Steps per Episode]{
    \label{steps_random}
    \framebox{\includegraphics[width=0.48\linewidth]{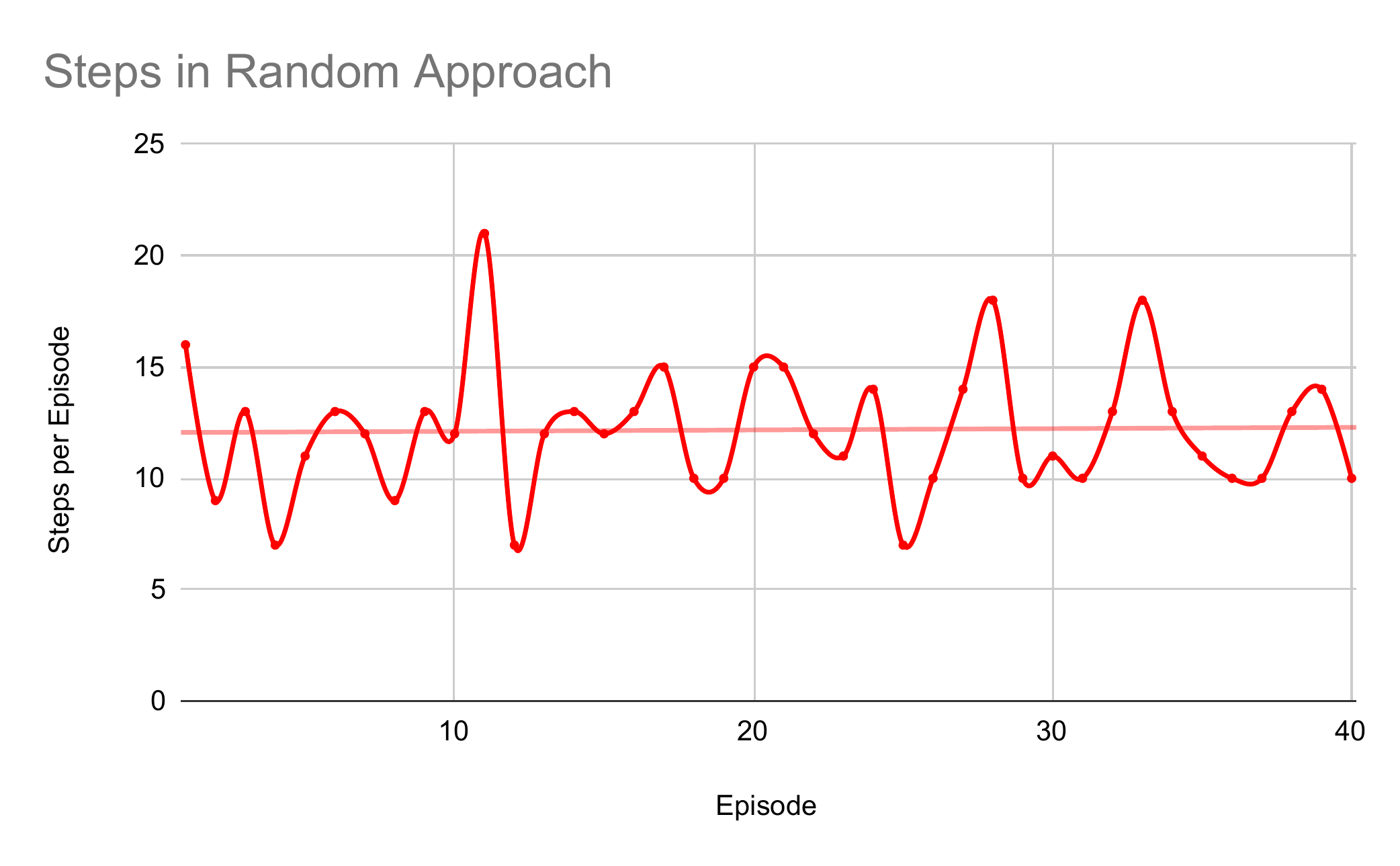}} } 
\subfloat[Final Workload Size per Episode]{
    \label{final_workload_random}
    \framebox{\includegraphics[width=0.48\linewidth]{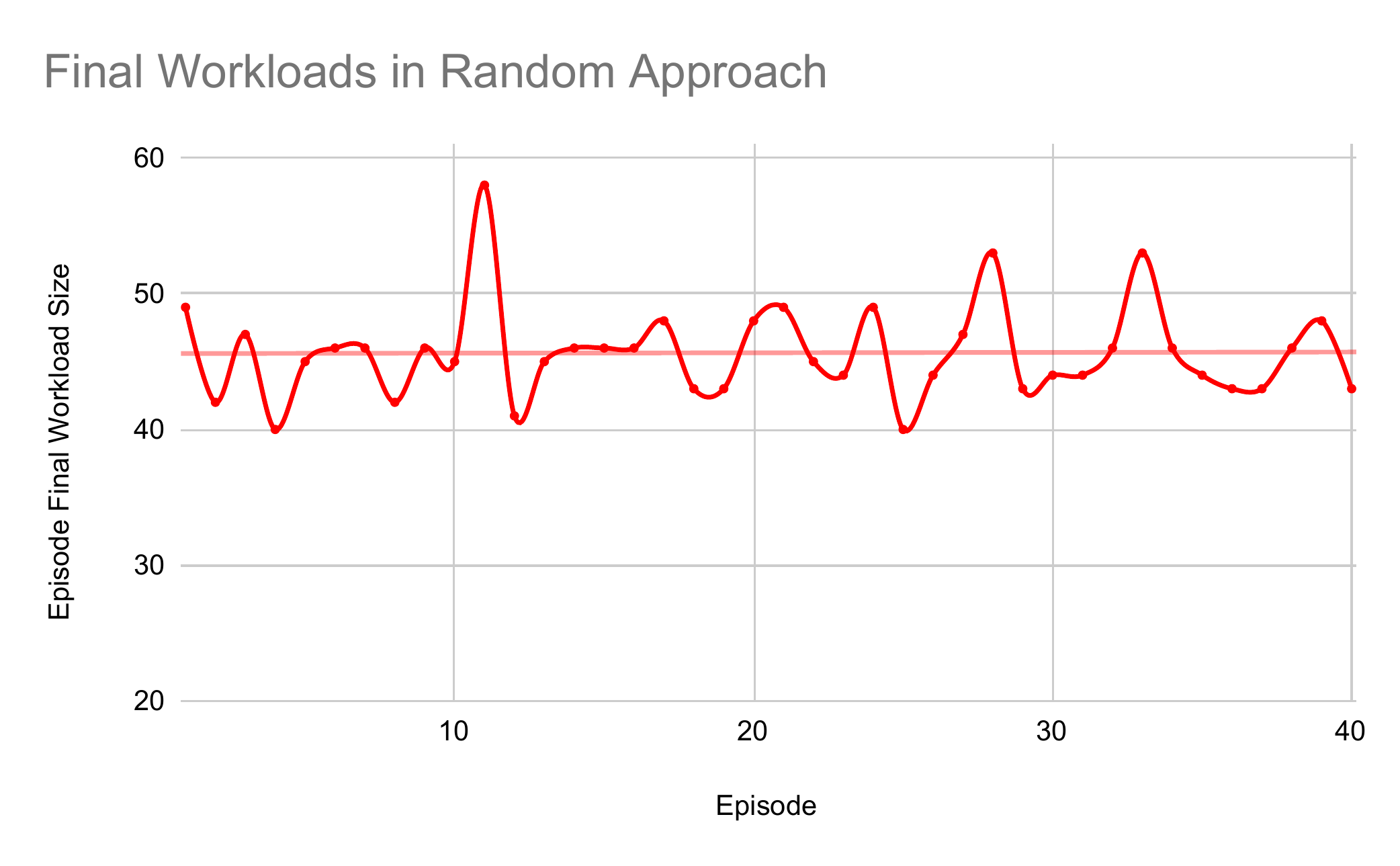}}}
\caption{Random Approach}
\end{figure}

The random approach was executed for 40 episodes.
Figure~\ref{steps_random} shows that in the random approach, the number of steps for increasing the workload is high. The trendline is between 10 and 15 steps, and it is constant (as expected from a random method). It indicates that on average, no extreme change happened overtime in the state of the system and that the SUT remains stable. \\

As shown in Figure~\ref{final_workload_random}, in the random approach, the size of the final workload, which hit the thresholds in each episode, is generally between 40 and 50. This size is smaller than the general size of the final workload in the baseline approach because here, the increments in the workload size were applied to just one transaction per step and not all of them.
Thus this approach could find smaller workloads that hit the threshold.% In this figure, the trend line is constant as it is expected from a random method.

\paragraph{Results of the Q-learning Approach} 

\begin{figure}[h]
\label{qlearning}
\centering
\subfloat[Number of Steps per Episode]{
    \label{steps_q-learning}
    \framebox{\includegraphics[width=0.48\linewidth]{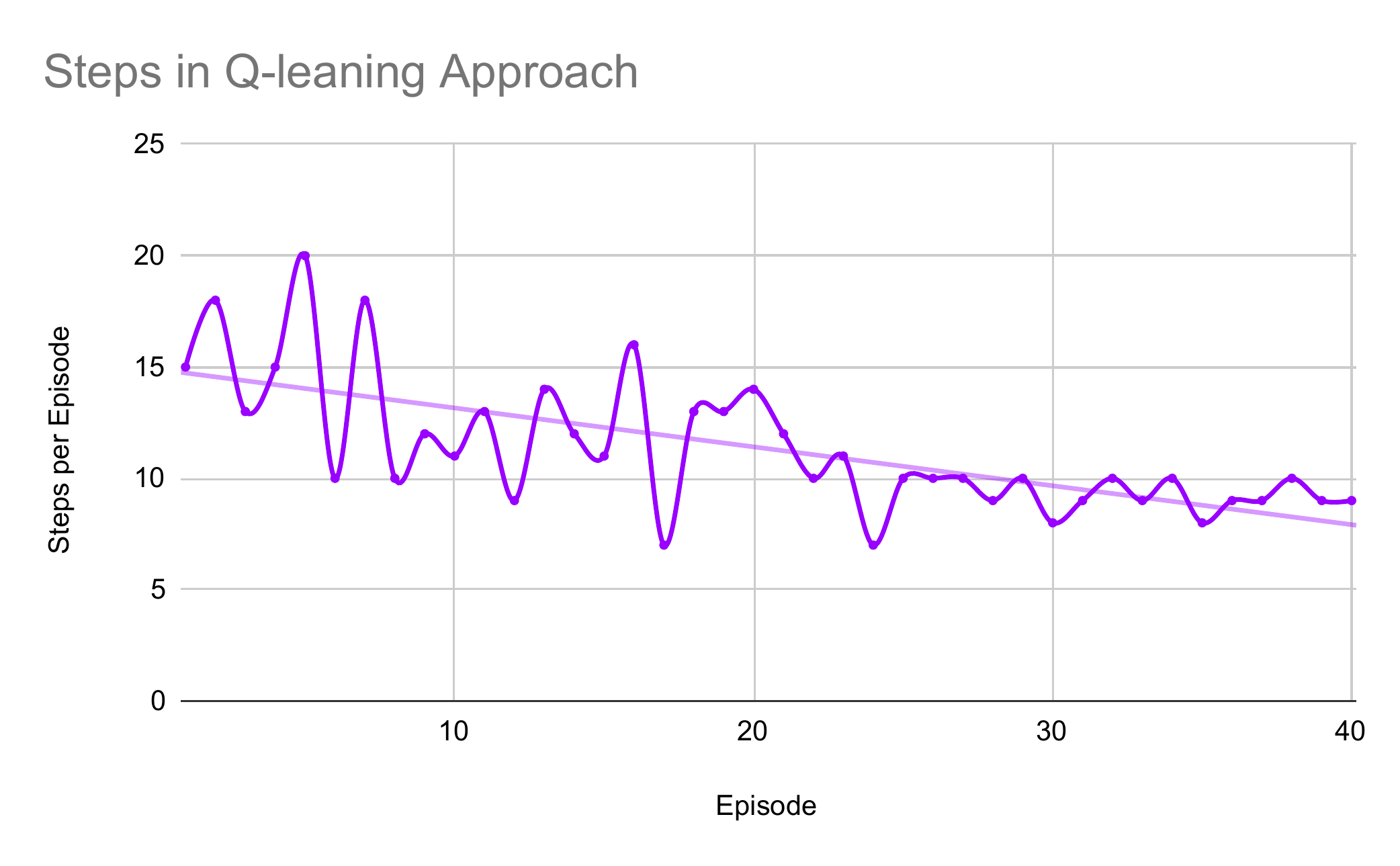}}} 
\subfloat[Final Workload Size per Episode]{
    \label{final_workload_qlearning}
    \framebox{\includegraphics[width=0.48\linewidth]{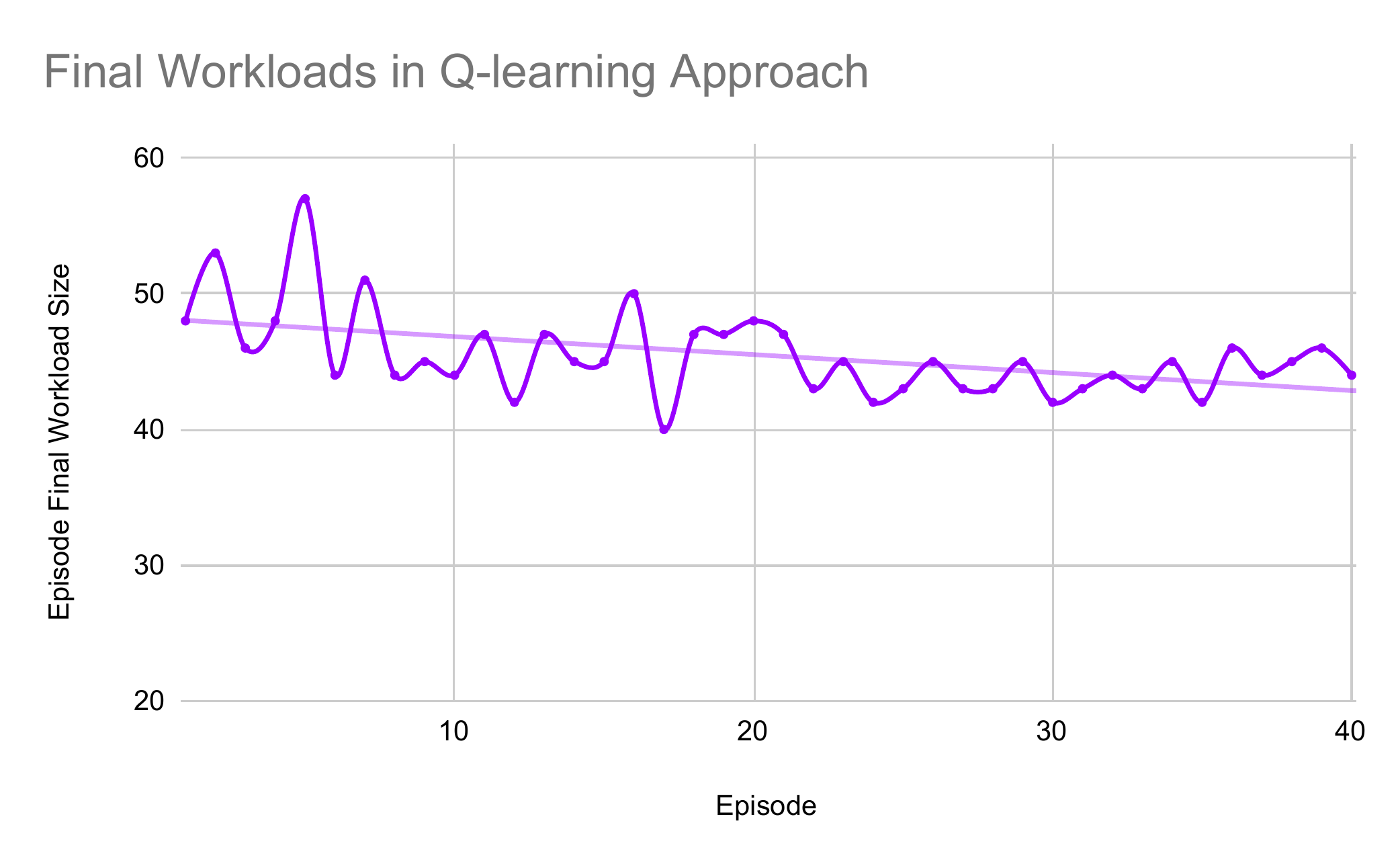}}}
\caption{Q-learning Approach}
\end{figure}

The q-learning approach was executed for 40 episodes.
Figure~\ref{steps_q-learning} shows that the number of steps in each episode is, on average, between 5 and 15 steps. 
We can see that the diversity of the number of steps is high in the first episodes, which is similar to the results of the random approach as shown in Figure~\ref{steps_random}.
But the diversity decreases over time, and after episode 25 i.e., the last 16 episodes, the number of steps converge and appear in the range of 8 to 10.
This indicates that the agent has developed a policy, and it is following it. The policy that has been learned and converged over time and is not changing drastically after episode 25.

Another practical factor of the convergence in this approach is because of using the decaying $\varepsilon$-greedy method. However, the q-learning method is expected to converge to the optimal policy with the probability of one \cite{10.1007/BF00992698}, using decaying $\varepsilon$-greedy can accelerate the convergence.
As mentioned in Section~\ref{qlearning_method}, at the beginning of the learning, the actions are chosen more randomly to allow the agent to explore different actions and learn the consequences of taking them (by receiving the reward from taking that action). But over the time, the probability of choosing random actions decreases, also the probability of choosing the best action due to the main policy increases (policy derived from the q-table). 
Therefore, in each step, the actions (i.e., the transactions that their workloads are incremented) are chosen more randomly at the beginning of the learning (first episodes), but they are chosen less randomly and more intelligently based on the learned policy in the last episodes. 
In the final episodes, the process of generating the workload is following the same policy which leads to convergence of the number of steps in each episode.\\

In addition, the trendline in Figure~\ref{steps_q-learning} decreases over the time, which means that the number of steps for generating a workload that hits the thresholds, is decreasing. This indicates that the agent has learned to take more efficient actions and to choose the best candidate transitions to increment its workload in each step. 
Choosing the best actions intelligently in the lasts episodes leads to generating an effective workload in fewer steps.
The convergence and degradation in the number of steps shows that the agent has found the optimal policy for generating effective workload which hits the threshold in fewer steps.\\

Figure~\ref{final_workload_qlearning} shows that in the q-learning approach, the size of the final workload in each episode is small, and it is, on average, between 40 and 50. 
The diversity of the final workload size is high in the first 20 episodes, but it decreases over time and converges to the range of 42 to 46 after episode 22 (i.e., last 19 episodes).
Also, the trendline shows, the final workload size decreases during the time.

\begin{figure}[h]
\label{qlearning}
\centering
\subfloat[Number of Steps per Episode]{
    \label{steps_dqn}
    \framebox{\includegraphics[width=0.48\linewidth]{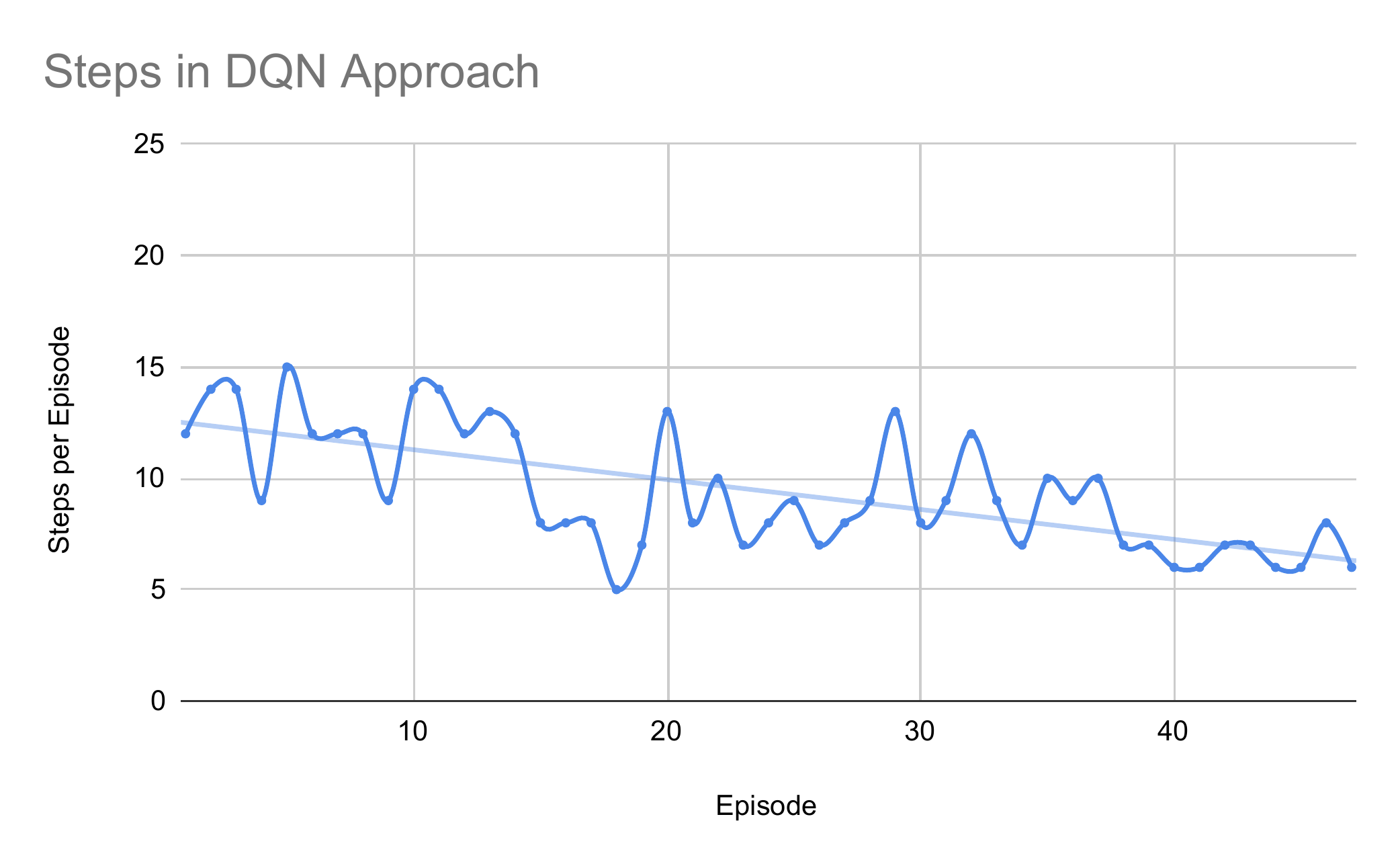}}} 
\subfloat[Final Workload Size per Episode]{
    \label{final_workload_dqn}
    \framebox{\includegraphics[width=0.48\linewidth]{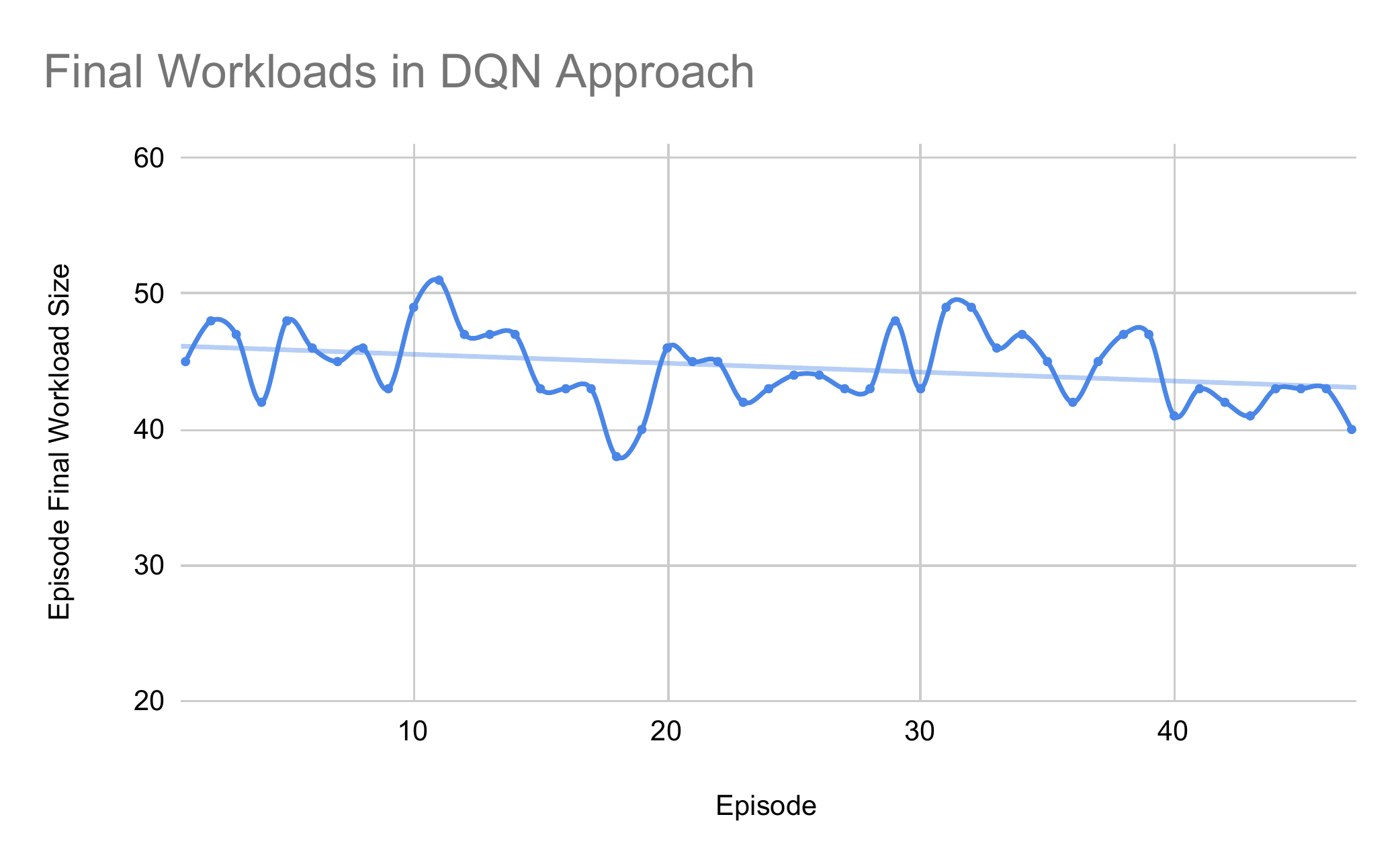}}}
\caption{DQN Approach}
\end{figure}

\paragraph{Results of the DQN Approach} 
The DQN approach was executed for 47 episodes.
As shown in Figure~\ref{steps_dqn}, The number of steps are between 5 and 15. 
After episode 38 i.e., the last 10 episodes, the number of steps are staying in the range of 6 to 8. 
Like the q-learning approach and because of the same reason (i.e., using decaying $\varepsilon$-greedy method in the policy), the number of steps are converging in the last episodes. 
The steps are staying in the range of 6 to 8 after episode 38. Also, the slope of the trend line show that the number of steps for generating the effective workload decreases over time. 
Based on the convergence and the decrease in the number of steps, we can see that the agent has learned the optimal policy, and it is following that policy.
We can also see that in comparison to the q-learning approach, the convergence is occurring after more episodes, but the convergence range is less than the convergence range in the q-learning approach.  \\

In Figure~\ref{final_workload_dqn}, the trendline shows that the final workload is generally between 40 and 50. 
We can see in the figure that after episode 40 i.e., the last 8 episodes, the size of the effective workload is altering in the range of 40 to 43. This shows that it is converging after episode 40, and the size of the final workload has reduced after this episode. Comparing to the q-learning method, the convergence and the decrease in the size of effective workload are happening in later episodes. But the convergence range is less than the range in the q-learning method (which is 8 to 10).

\newpage
\section{Discussion}
\label{sec:discussion}

%\instructions{Here you present an interpretation of the results and assess their significance. Discuss the possible consequences of the results, and present any recommendations. It is important that you report on whether you have achieved the goals you set and thus answered your question and achieved the purpose of the work. The section should also contain reflections on the work, such as its limitations. You can also discuss solutions to problems that you have identified and discussed previously, or address other problems that the work has not addressed, questions that have not been answered. Also link your results to previous work. This way, the discussion can become a conversation with what you wrote in previous sections. Finally, put your own work into a wider context, broaden your perspective. Can your results be generalized? Can what you have done be used in some other context?}\\

In this section, we highlight the answers to our research questions. We review the solutions and discuss the results (important bits in \textit{italic}). We also discuss the threats to the validity of our results.

\paragraph{RQ1.} Answering RQ1 required formulating an RL solution for the test load generation problem. In order to apply RL based solution, a mapping of the real-world problem into an RL problem is a pre-requisite.
We formulated and presented our mapping in Section~\ref{sec:approach}, and provided our RL approach for load test generation in detail.
We defined the agent, environment, and the learning principles including states, actions, observations, and reward function in our RL problem. 
\textit{As discussed, we considered the SUT as the environment. We declared each state base on the last average error rate and average response time observed from the SUT.
The actions were defined as choosing a transaction, increasing its workload, and applying it to the SUT. 
The average error rate and average response time were the agent's observations of the SUT. And the reward was calculated according to the observations of the SUT.
We also chose two different RL methods for in our approach namely q-learning and DQN.}

\paragraph{RQ2.} To demonstrate the applicability of our proposed RL-based load testing approaches, which is RQ2, we implemented and applied the q-learning and DQN approaches on an SUT in Section~\ref{sec:evaluation}. We set up an e-commerce store on a local server using open-source and heavily maintained CMS and plugin i.e. WordPress and the WooCommerce plugin. We used JMeter for generating the load scenarios (produced by the RL agent), applying them on the SUT, and recording the error rate and response time. \textit{Results show that RL-based test load generation approaches are applicable for performance testing of real-world applications.}

\paragraph{RQ3.} To answer the third research question, we conducted an experiment. In the experiment, we executed four approaches (treatments) for load test generation, which were a baseline, a random, and the proposed q-learning and DQN approaches. The results of the experiment are provided in Section \ref{sec:results}.
Results from our experiment show that the baseline approach (in which the workload was increased per all transactions in each step) generally produced a larger effective workload. \textit{Thus we can conclude that, in comparison to other approaches, the baseline approach for workload generation is not efficient in terms of the size of generated workload.}

In contrast, the random approach did perform better than the baseline approach. 
%However, the results indicate no extreme change in the observed execution environment (response rate) of the SUT.
\textit{In comparison with RL-based approaches, the random approach was generally not attaining an optimal workload size, and the diversity of the applied workload sizes remained high.}

The results show that the number of steps per episode and the size of the effective workload in each episode of both of the RL approaches are converging to a lower number in the last episodes.  Results indicate that the \textit{q-learning approach converges faster (in terms of number of steps and optimal workload size) than the DQN approach.} This is expected since the q-learning approach only has six states while the DQN approach has much more states (\textit{error\_rate\_threshold}$\times$\textit{response\_time\_threshold} states).

Findings also revealed that the DQN approach converges to lower values in both metrics. Meaning that \textit{in comparison to q-learning-based approach, the DQN approach took more time to converge. However, the DQN was more efficient in terms of finding the optimal workload sizes.} Based on our results, we believe that the DQN approach can perform even better after extensive tuning of the hyper-parameters.
%The benefit of DQN over q-learning is the states are much more, and the things I mentioned before and that we don't need to define points which separate states where choosing these points to define the sates is dependent on the situation and affects the efficiency of learning.\\
%We executed the DQN method with different combination values for the hyper-parameters. But still, we may be able to get better results from the DQN method by testing it with other combination of the hyper-parameters and finding the best configuration.\\
\textit{Finally, we can conclude that both of the proposed RL approaches for load test generation converges to the optimal policy and performs better than the baseline and random approach.}\\

\subsection{Threats to Validity}
%\todoB{Add the items in the \cite{ahmad2019exploratory} paper }\\

%\todoB{Write this section based on the threats to validity section checklist in section 8.8 and 8.9 of the experimentation in software engineering book}

Load scenario generation is heavily dependent on the hardware of the SUT and its execution environment. External factors (such as running the SUT on a shared hosting server) might alter the results. In this section, we address the potential threats to validity based on the classification presented by Runeson and H\"ost~\cite{RuHo09}.
%cite the guidlines for casestudy
%\todoB{ Using a single subject in the experiment e.g. one SUT can be considered as a generalization validity threat.}\\

\bigskip
Construct validity: This aspect of validity reflects to what extent the studied operational
measures really represent what the researcher has in mind. A misunderstood question is an example of a potential construct validity threat. To tackle potential threats to our construct validity, we have benefited the guides from multiple researchers in the problem formulation and used well-established guidelines for conducting our study.\\
%to review

Internal validity: This aspect of validity is of concern with the validity and credibility of the obtained results. We tackle the potential threats to internal validity by executing the experiment on a dedicated local server with no other processes running on it. However, there were several uncontrolled factors (such as the operating system's processes) that might have affected our results.\\

External validity: This aspect of validity is concerned with to what extent it is possible to generalize the findings, and to what extent the findings are of interest to other people outside the investigated case. In our case, the results are obtained by executing the experiment on one SUT, and we, therefore, do not claim that the results can be generalized to other cases. However, we chose open-source and heavily maintained SUT, so that the results can be generalized to other similar cases (WooCommerce-based e-commerce applications). In addition, our results can also be of interest to performance testing researchers and practitioners working with load testing via JMeter.

\bigskip
Reliability: This aspect is concerned with to what extent the data and the analysis are
dependent on the specific researchers. Hypothetically, if another researcher, later on,
conducted the same study, the result should be the same. Threats to this aspect of
validity are tackled by receiving feedback from multiple researchers in experiment planning and execution. In addition, we provided enough details on our experiment setup for replication.

\newpage
\section{Conclusions}
\label{sec:conclusions}

One critical activity in performance testing is the generation of load tests. Existing load testing approaches heavily rely on the system models or source code.  This thesis aims to propose and evaluate a model-free and an intelligent approach for load test generation.

%In this thesis, we proposed an approach for load test generation. We presented a mapping between the load test generation and reinforcement learning and formulated the problem as a reinforcement learning problem. The approach addresses the problem of load test generation by introducing an intelligent adaptive load tester that satisfies the load testing objectives using reinforcement learning. The proposed approach can be executed without having access to the source code and system model of the application under test in different execution conditions. 
%We present the two variant architectures of our approach and the learning procedure for generating load tests. We demonstrated the applicability of our approach by applying it to a real-world software system. In addition, we also conducted an experiment for comparing the two variants of our approach with a random and baseline load test generators.

In this thesis, we formulated the problem of efficient test generation for load testing into an RL problem. We presented an RL-driven model-free approach for generating effective workloads in performance testing. We mapped the real-world problem of test load generation into the RL context and discussed our approach in detail. We evaluate the applicability of our proposed RL based test load generation on a real-world software system.
In addition, we conducted an experiment to compare the efficiency of two different proposed RL-based, q-learning, and DQN methods for effective workload generation.
For the experiment, we implemented our proposed approach and prepared the requirements for testing it.
We set up an e-commerce store on a local server using WordPress and its WooComerce plugin. Then we applied the workloads generated by our approach to it. 
The workloads consisted of a different number of various transactions (operations) generated using JMeter. 
We performed a one factor-four treatment experiment on the same SUT where the factor for the experiment was 
"test load generation method" with treatments of
baseline method, a random method, the q-learning method, and the DQN method. 
We executed each treatment for around 40 episodes. Each episode contained several load generation steps, and it would be finished when the generated workload produced an error rate or response time larger than the defined threshold.

The results indicated that, in general, the baseline approach was not efficient in terms of the size of the generated effective workload per episode.
In the random approach, the average size of generated effective workload was smaller than the baseline approach, which means that the random approach performed better than the baseline approach in test load generation.
In addition, the results showed that both of the RL-based methods performed better than the random and baseline approaches. The results show that effective workload size and the number of steps taken for generating workload in each episode converges to a lower value in both q-learning and DQN approaches. 
The q-learning approach converged faster than the DQN. However, the DQN approach converged to lower values for the workload sizes.
We can conclude from the results that both of the proposed RL approaches learned an optimal policy to generate optimal workloads efficiently.

The RL-based approaches performed batter in our experiment and do not require access to the system models or source code. In addition, the learned policy can be reused in further similar situations (stages) of testing, e.g., regression testing and in the process of Development and Operations (DevOps) incremental release testing.

In the future, we plan to extend our approach to support performance testing for software product lines. In software product lines, the derived products are variants of the standard product, and the learned policy for load generation can be reused on multiple derived products. This can significantly reduce the performance testing time for the software product line.
\newpage

%\input{./Acknowledgments}

% ============================= References ============================
\newpage
\bibliographystyle{IEEEtran}
\bibliography{main}

% Generated by IEEEtran.bst, version: 1.14 (2015/08/26)
\begin{thebibliography}{10}
\providecommand{\url}[1]{#1}
\csname url@samestyle\endcsname
\providecommand{\newblock}{\relax}
\providecommand{\bibinfo}[2]{#2}
\providecommand{\BIBentrySTDinterwordspacing}{\spaceskip=0pt\relax}
\providecommand{\BIBentryALTinterwordstretchfactor}{4}
\providecommand{\BIBentryALTinterwordspacing}{\spaceskip=\fontdimen2\font plus
\BIBentryALTinterwordstretchfactor\fontdimen3\font minus
  \fontdimen4\font\relax}
\providecommand{\BIBforeignlanguage}[2]{{%
\expandafter\ifx\csname l@#1\endcsname\relax
\typeout{** WARNING: IEEEtran.bst: No hyphenation pattern has been}%
\typeout{** loaded for the language `#1'. Using the pattern for}%
\typeout{** the default language instead.}%
\else
\language=\csname l@#1\endcsname
\fi
#2}}
\providecommand{\BIBdecl}{\relax}
\BIBdecl

\bibitem{mm@web}
\BIBentryALTinterwordspacing
G.~Linden. (2006) Marissa mayer at web 2.0. [Online]. Available:
  \url{http://glinden.blogspot.com/2006/11/marissa-mayer-at-web-20.html}
\BIBentrySTDinterwordspacing

\bibitem{glindenmakedatausefull}
\BIBentryALTinterwordspacing
------. (2006) Make your data useful. [Online]. Available:
  \url{http://sites.google.com/site/glinden/Home/StanfordDataMining.2006-11-29.ppt}
\BIBentrySTDinterwordspacing

\bibitem{jin2012understanding}
G.~Jin, L.~Song, X.~Shi, J.~Scherpelz, and S.~Lu, ``Understanding and detecting
  real-world performance bugs,'' \emph{ACM SIGPLAN Notices}, vol.~47, no.~6,
  pp. 77--88, 2012.

\bibitem{weyuker2000experience}
E.~J. Weyuker and F.~I. Vokolos, ``Experience with performance testing of
  software systems: issues, an approach, and case study,'' \emph{IEEE
  transactions on software engineering}, vol.~26, no.~12, pp. 1147--1156, 2000.

\bibitem{lavenberg1983computer}
S.~Lavenberg, \emph{Computer performance modeling handbook}.\hskip 1em plus
  0.5em minus 0.4em\relax Elsevier, 1983.

\bibitem{zhang2011automatic}
P.~Zhang, S.~Elbaum, and M.~B. Dwyer, ``Automatic generation of load tests,''
  in \emph{Proceedings of the 2011 26th IEEE/ACM International Conference on
  Automated Software Engineering}.\hskip 1em plus 0.5em minus 0.4em\relax IEEE
  Computer Society, 2011, pp. 43--52.

\bibitem{zhang2002automated}
J.~Zhang and S.~C. Cheung, ``Automated test case generation for the stress
  testing of multimedia systems,'' \emph{Software: Practice and Experience},
  vol.~32, no.~15, pp. 1411--1435, 2002.

\bibitem{syer2011identifying}
M.~D. Syer, B.~Adams, and A.~E. Hassan, ``Identifying performance deviations in
  thread pools,'' in \emph{2011 27th IEEE International Conference on Software
  Maintenance (ICSM)}.\hskip 1em plus 0.5em minus 0.4em\relax IEEE, 2011, pp.
  83--92.

\bibitem{koo2019pyse}
J.~Koo, C.~Saumya, M.~Kulkarni, and S.~Bagchi, ``Pyse: Automatic worst-case
  test generation by reinforcement learning,'' in \emph{2019 12th IEEE
  Conference on Software Testing, Validation and Verification (ICST)}.\hskip
  1em plus 0.5em minus 0.4em\relax IEEE, 2019, pp. 136--147.

\bibitem{sutton1998introduction}
R.~S. Sutton, A.~G. Barto \emph{et~al.}, \emph{Introduction to reinforcement
  learning}.\hskip 1em plus 0.5em minus 0.4em\relax MIT press Cambridge, 1998,
  vol. 135.

\bibitem{ibidunmoye2015performance}
O.~Ibidunmoye, F.~Hern{\'a}ndez-Rodriguez, and E.~Elmroth, ``Performance
  anomaly detection and bottleneck identification,'' \emph{ACM Computing
  Surveys (CSUR)}, vol.~48, no.~1, p.~4, 2015.

\bibitem{chandola2009anomaly}
V.~Chandola, A.~Banerjee, and V.~Kumar, ``Anomaly detection: A survey,''
  \emph{ACM computing surveys (CSUR)}, vol.~41, no.~3, p.~15, 2009.

\bibitem{ISO/IEC}
{ISO 25000}, ``{ISO/IEC 25010 - System and software quality models},'' 2019,
  available at
  {\url{https://iso25000.com/index.php/en/iso-25000-standards/iso-25010}},
  Retrieved July, 2019.

\bibitem{glinz2007non}
M.~Glinz, ``On non-functional requirements,'' in \emph{15th IEEE International
  Requirements Engineering Conference (RE 2007)}.\hskip 1em plus 0.5em minus
  0.4em\relax IEEE, 2007, pp. 21--26.

\bibitem{chung2012non}
L.~Chung, B.~A. Nixon, E.~Yu, and J.~Mylopoulos, \emph{Non-functional
  requirements in software engineering}.\hskip 1em plus 0.5em minus 0.4em\relax
  Springer Science \& Business Media, 2012, vol.~5.

\bibitem{cortellessa2011model}
V.~Cortellessa, A.~Di~Marco, and P.~Inverardi, \emph{Model-based software
  performance analysis}.\hskip 1em plus 0.5em minus 0.4em\relax Springer
  Science \& Business Media, 2011.

\bibitem{harchol2013performance}
M.~Harchol-Balter, \emph{Performance modeling and design of computer systems:
  queueing theory in action}.\hskip 1em plus 0.5em minus 0.4em\relax Cambridge
  University Press, 2013.

\bibitem{kant1992introduction}
K.~Kant and M.~Srinivasan, \emph{Introduction to computer system performance
  evaluation}.\hskip 1em plus 0.5em minus 0.4em\relax McGraw-Hill College,
  1992.

\bibitem{10.5555/574566}
A.~Geraci, F.~Katki, L.~McMonegal, B.~Meyer, J.~Lane, P.~Wilson, J.~Radatz,
  M.~Yee, H.~Porteous, and F.~Springsteel, \emph{IEEE Standard Computer
  Dictionary: Compilation of IEEE Standard Computer Glossaries}.\hskip 1em plus
  0.5em minus 0.4em\relax IEEE Press, 1991.

\bibitem{jiang2015survey}
Z.~M. Jiang and A.~E. Hassan, ``A survey on load testing of large-scale
  software systems,'' \emph{IEEE Transactions on Software Engineering},
  vol.~41, no.~11, pp. 1091--1118, 2015.

\bibitem{gregg2013systems}
B.~Gregg, \emph{Systems performance: enterprise and the cloud}.\hskip 1em plus
  0.5em minus 0.4em\relax Pearson Education, 2013.

\bibitem{beizer1984software}
B.~Beizer, \emph{Software system testing and quality assurance}.\hskip 1em plus
  0.5em minus 0.4em\relax Van Nostrand Reinhold Co., 1984.

\bibitem{10.5555/541177}
T.~M. Mitchell, \emph{Machine Learning}, 1st~ed.\hskip 1em plus 0.5em minus
  0.4em\relax USA: McGraw-Hill, Inc., 1997.

\bibitem{lin1992self}
L.-J. Lin, ``Self-improving reactive agents based on reinforcement learning,
  planning and teaching,'' \emph{Machine learning}, vol.~8, no. 3-4, pp.
  293--321, 1992.

\bibitem{mnih2015human}
V.~Mnih, K.~Kavukcuoglu, D.~Silver, A.~A. Rusu, J.~Veness, M.~G. Bellemare,
  A.~Graves, M.~Riedmiller, A.~K. Fidjeland, G.~Ostrovski \emph{et~al.},
  ``Human-level control through deep reinforcement learning,'' \emph{Nature},
  vol. 518, no. 7540, pp. 529--533, 2015.

\bibitem{schaul2015prioritized}
T.~Schaul, J.~Quan, I.~Antonoglou, and D.~Silver, ``Prioritized experience
  replay,'' \emph{arXiv preprint arXiv:1511.05952}, 2015.

\bibitem{DifBatch}
\BIBentryALTinterwordspacing
J.~Brownlee. (2018) Difference between a batch and an epoch in a neural
  network. [Online]. Available:
  \url{https://machinelearningmastery.com/difference-between-a-batch-and-an-epoch/}
\BIBentrySTDinterwordspacing

\bibitem{hasselt2010double}
H.~V. Hasselt, ``Double q-learning,'' in \emph{Advances in neural information
  processing systems}, 2010, pp. 2613--2621.

\bibitem{van2016deep}
H.~Van~Hasselt, A.~Guez, and D.~Silver, ``Deep reinforcement learning with
  double q-learning,'' in \emph{Thirtieth AAAI conference on artificial
  intelligence}, 2016.

\bibitem{menasce2002load}
D.~A. Menasc{\'e}, ``Load testing, benchmarking, and application performance
  management for the web,'' in \emph{Int. CMG Conference}, 2002, pp. 271--282.

\bibitem{apte2017autoperf}
V.~Apte, T.~Viswanath, D.~Gawali, A.~Kommireddy, and A.~Gupta, ``Autoperf:
  Automated load testing and resource usage profiling of multi-tier internet
  applications,'' in \emph{Proceedings of the 8th ACM/SPEC on International
  Conference on Performance Engineering}.\hskip 1em plus 0.5em minus
  0.4em\relax ACM, 2017, pp. 115--126.

\bibitem{jindal2019performance}
A.~Jindal, V.~Podolskiy, and M.~Gerndt, ``Performance modeling for cloud
  microservice applications,'' in \emph{Proceedings of the 2019 ACM/SPEC
  International Conference on Performance Engineering}.\hskip 1em plus 0.5em
  minus 0.4em\relax ACM, 2019, pp. 25--32.

\bibitem{briand2005stress}
L.~C. Briand, Y.~Labiche, and M.~Shousha, ``Stress testing real-time systems
  with genetic algorithms,'' in \emph{Proceedings of the 7th annual conference
  on Genetic and evolutionary computation}.\hskip 1em plus 0.5em minus
  0.4em\relax ACM, 2005, pp. 1021--1028.

\bibitem{ayala2018one}
V.~Ayala-Rivera, M.~Kaczmarski, J.~Murphy, A.~Darisa, and A.~O.
  Portillo-Dominguez, ``One size does not fit all: In-test workload adaptation
  for performance testing of enterprise applications,'' in \emph{Proceedings of
  the 2018 ACM/SPEC International Conference on Performance Engineering}.\hskip
  1em plus 0.5em minus 0.4em\relax ACM, 2018, pp. 211--222.

\bibitem{gu2009search}
Y.~Gu and Y.~Ge, ``Search-based performance testing of applications with
  composite services,'' in \emph{2009 International Conference on Web
  Information Systems and Mining}.\hskip 1em plus 0.5em minus 0.4em\relax IEEE,
  2009, pp. 320--324.

\bibitem{di2007search}
M.~Di~Penta, G.~Canfora, G.~Esposito, V.~Mazza, and M.~Bruno, ``Search-based
  testing of service level agreements,'' in \emph{Proceedings of the 9th annual
  conference on Genetic and evolutionary computation}.\hskip 1em plus 0.5em
  minus 0.4em\relax ACM, 2007, pp. 1090--1097.

\bibitem{garousi2010genetic}
V.~Garousi, ``A genetic algorithm-based stress test requirements generator tool
  and its empirical evaluation,'' \emph{IEEE Transactions on Software
  Engineering}, vol.~36, no.~6, pp. 778--797, 2010.

\bibitem{garousi2008traffic}
V.~Garousi, L.~C. Briand, and Y.~Labiche, ``Traffic-aware stress testing of
  distributed real-time systems based on uml models using genetic algorithms,''
  \emph{Journal of Systems and Software}, vol.~81, no.~2, pp. 161--185, 2008.

\bibitem{yang1996towards}
C.-S.~D. Yang and L.~L. Pollock, ``Towards a structural load testing tool,'' in
  \emph{ACM SIGSOFT Software Engineering Notes}, vol.~21, no.~3.\hskip 1em plus
  0.5em minus 0.4em\relax ACM, 1996, pp. 201--208.

\bibitem{draheim2006realistic}
D.~Draheim, J.~Grundy, J.~Hosking, C.~Lutteroth, and G.~Weber, ``Realistic load
  testing of web applications,'' in \emph{Conference on Software Maintenance
  and Reengineering (CSMR'06)}.\hskip 1em plus 0.5em minus 0.4em\relax IEEE,
  2006, pp. 11--pp.

\bibitem{lutteroth2008modeling}
C.~Lutteroth and G.~Weber, ``Modeling a realistic workload for performance
  testing,'' in \emph{2008 12th International IEEE Enterprise Distributed
  Object Computing Conference}.\hskip 1em plus 0.5em minus 0.4em\relax IEEE,
  2008, pp. 149--158.

\bibitem{shams2006model}
M.~Shams, D.~Krishnamurthy, and B.~Far, ``A model-based approach for testing
  the performance of web applications,'' in \emph{Proceedings of the 3rd
  international workshop on Software quality assurance}.\hskip 1em plus 0.5em
  minus 0.4em\relax ACM, 2006, pp. 54--61.

\bibitem{vogele2018wessbas}
C.~V{\"o}gele, A.~van Hoorn, E.~Schulz, W.~Hasselbring, and H.~Krcmar,
  ``Wessbas: extraction of probabilistic workload specifications for load
  testing and performance prediction—a model-driven approach for
  session-based application systems,'' \emph{Software \& Systems Modeling},
  vol.~17, no.~2, pp. 443--477, 2018.

\bibitem{ferme2018declarative}
V.~Ferme and C.~Pautasso, ``A declarative approach for performance tests
  execution in continuous software development environments,'' in
  \emph{Proceedings of the 2018 ACM/SPEC International Conference on
  Performance Engineering}.\hskip 1em plus 0.5em minus 0.4em\relax ACM, 2018,
  pp. 261--272.

\bibitem{ferme2017towards}
------, ``Towards holistic continuous software performance assessment,'' in
  \emph{Proceedings of the 8th ACM/SPEC on International Conference on
  Performance Engineering Companion}.\hskip 1em plus 0.5em minus 0.4em\relax
  ACM, 2017, pp. 159--164.

\bibitem{schulz2019behavior}
H.~Schulz, D.~Okanovi{\'c}, A.~van Hoorn, V.~Ferme, and C.~Pautasso,
  ``Behavior-driven load testing using contextual knowledge-approach and
  experiences,'' in \emph{Proceedings of the 2019 ACM/SPEC International
  Conference on Performance Engineering}.\hskip 1em plus 0.5em minus
  0.4em\relax ACM, 2019, pp. 265--272.

\bibitem{malik2013automatic}
H.~Malik, H.~Hemmati, and A.~E. Hassan, ``Automatic detection of performance
  deviations in the load testing of large scale systems,'' in \emph{Proceedings
  of the 2013 International Conference on Software Engineering}.\hskip 1em plus
  0.5em minus 0.4em\relax IEEE Press, 2013, pp. 1012--1021.

\bibitem{grechanik2012automatically}
M.~Grechanik, C.~Fu, and Q.~Xie, ``Automatically finding performance problems
  with feedback-directed learning software testing,'' in \emph{2012 34th
  International Conference on Software Engineering (ICSE)}.\hskip 1em plus
  0.5em minus 0.4em\relax IEEE, 2012, pp. 156--166.

\bibitem{ahmad2019exploratory}
T.~Ahmad, A.~Ashraf, D.~Truscan, and I.~Porres, ``Exploratory performance
  testing using reinforcement learning,'' in \emph{2019 45th Euromicro
  Conference on Software Engineering and Advanced Applications (SEAA)}.\hskip
  1em plus 0.5em minus 0.4em\relax IEEE, 2019, pp. 156--163.

\bibitem{basili1988tame}
V.~R. Basili and H.~D. Rombach, ``The tame project: Towards
  improvement-oriented software environments,'' \emph{IEEE Transactions on
  software engineering}, vol.~14, no.~6, pp. 758--773, 1988.

\bibitem{holz2006research}
H.~J. Holz, A.~Applin, B.~Haberman, D.~Joyce, H.~Purchase, and C.~Reed,
  ``Research methods in computing: What are they, and how should we teach
  them?'' in \emph{Working group reports on ITiCSE on Innovation and technology
  in computer science education}, 2006, pp. 96--114.

\bibitem{10.5555/2349018}
C.~Wohlin, P.~Runeson, M.~Hst, M.~C. Ohlsson, B.~Regnell, and A.~Wessln,
  \emph{Experimentation in Software Engineering}.\hskip 1em plus 0.5em minus
  0.4em\relax Springer Publishing Company, Incorporated, 2012.

\bibitem{rl4j}
R.~Fiszel, ``Rl4j: Reinforcement learning for java,''
  \url{https://github.com/eclipse/deeplearning4j/tree/master/rl4j}, accessed:
  2020-02-01.

\bibitem{Deeplearning4j}
``Deeplearning4j: Open-source distributed deep learning for the jvm, apache
  software foundation license 2.0,'' \url{https://deeplearning4j.org},
  accessed: 2020-04-21.

\bibitem{raj2019java}
R.~Raj, \emph{Java Deep Learning Cookbook}.\hskip 1em plus 0.5em minus
  0.4em\relax Packt Publishing Ltd, 2019.

\bibitem{ReinfLearnPlayPixels}
\BIBentryALTinterwordspacing
R.~Fiszel. (2016) Reinforcement learning and dqn, learning to play from pixels.
  [Online]. Available:
  \url{https://rubenfiszel.github.io/posts/rl4j/2016-08-24-Reinforcement-Learning-and-DQN.html}
\BIBentrySTDinterwordspacing

\bibitem{10.1145/1273463.1273476}
\BIBentryALTinterwordspacing
I.~Ciupa, A.~Leitner, M.~Oriol, and B.~Meyer, ``Experimental assessment of
  random testing for object-oriented software,'' in \emph{Proceedings of the
  2007 International Symposium on Software Testing and Analysis}, ser. ISSTA
  ’07.\hskip 1em plus 0.5em minus 0.4em\relax New York, NY, USA: Association
  for Computing Machinery, 2007, p. 84–94. [Online]. Available:
  \url{https://doi.org/10.1145/1273463.1273476}
\BIBentrySTDinterwordspacing

\bibitem{10.1109/TSE.1984.5010257}
\BIBentryALTinterwordspacing
J.~W. Duran and S.~C. Ntafos, ``An evaluation of random testing,'' \emph{IEEE
  Trans. Softw. Eng.}, vol.~10, no.~4, p. 438–444, Jul. 1984. [Online].
  Available: \url{https://doi.org/10.1109/TSE.1984.5010257}
\BIBentrySTDinterwordspacing

\bibitem{10.1007/BF00992698}
\BIBentryALTinterwordspacing
C.~J. C.~H. Watkins and P.~Dayan, ``Technical note: \cal q -learning,''
  \emph{Mach. Learn.}, vol.~8, no. 3–4, p. 279–292, May 1992. [Online].
  Available: \url{https://doi.org/10.1007/BF00992698}
\BIBentrySTDinterwordspacing

\bibitem{RuHo09}
\BIBentryALTinterwordspacing
P.~Runeson and M.~Höst, ``Guidelines for conducting and reporting case study
  research in software engineering,'' \emph{Empirical Software Engineering},
  vol.~14, no.~2, pp. 131--164, April 2009. [Online]. Available:
  \url{http://dx.doi.org/10.1007/s10664-008-9102-8}
\BIBentrySTDinterwordspacing

\end{thebibliography}
\addcontentsline{toc}{section}{References}

% ============================ Appendices =============================
\newpage
\begin{appendices}

	%\input{./Appendices/appendices1}
	%\clearpage
	
	%\input{./Appendices/appendices2}
	%\clearpage
	
\end{appendices}

\end{document}